\documentclass[a4paper,11pt]{article}

\usepackage{natbib}
\usepackage{eqnarray,amsmath}
\usepackage{amssymb}
\usepackage{graphicx}
\usepackage{epstopdf}
\usepackage{authblk}
\usepackage{xr}
\usepackage[linesnumbered]{algorithm2e}
\usepackage{subfigure}
\usepackage{bm}
\usepackage[margin=3cm]{geometry}
\usepackage[toc,page]{appendix}

\newcommand{\vect}[1]{\boldsymbol{#1}}

\makeatletter
\newcommand*{\addFileDependency}[1]{% argument=file name and extension
	\typeout{(#1)}
	\@addtofilelist{#1}
	\IfFileExists{#1}{}{\typeout{No file #1.}}
}
\makeatother
\newcommand*{\myexternaldocument}[1]{%
	\externaldocument{#1}%
	\addFileDependency{#1.tex}%
	\addFileDependency{#1.aux}%
}
\myexternaldocument{appendices}

\title{Robust Bayesian Synthetic Likelihood via a Semi-Parametric Approach}
\author[$\star$,$\dagger$]{Ziwen An\footnote{email: ziwen.an@hdr.qut.edu.au, ORCID ID: 0000-0002-9947-5182}}
\author[$\ddagger$,$\S$]{David J. Nott} 
\author[$\star$,$\dagger$]{Christopher Drovandi}
\affil[$\star$]{School of Mathematical Sciences, Queensland University of Technology, Australia}
\affil[ ]{}
\affil[$\dagger$]{Australian Research Council Centre of Excellence for Mathematical and Statistics Frontiers}
\affil[ ]{}
\affil[$\ddagger$]{Department of Statistics and Applied Probability, National University of Singapore}
\affil[ ]{}
\affil[$\S$]{Insitute of Operations Research and Analytics, National University of	Singapore, Singapore 117602}
\affil[ ]{}

\begin{document}
	
	\setlength{\parindent}{0pc}
	\setlength{\parskip}{1ex}
	
	\maketitle
	
	\begin{abstract}
		Bayesian synthetic likelihood (BSL) is now a well established method for performing approximate Bayesian parameter estimation for simulation-based models that do not possess a tractable likelihood function.  BSL approximates an intractable likelihood function of a carefully chosen summary statistic at a parameter value with a multivariate normal distribution.  The mean and covariance matrix of this normal distribution are estimated from independent simulations of the model.  Due to the parametric assumption implicit in BSL, it can be preferred to its nonparametric competitor, approximate Bayesian computation, in certain applications where a high-dimensional summary statistic is of interest. However,  despite several successful applications of BSL, its widespread use in scientific fields may be hindered by the strong normality assumption.  In this paper, we develop a semi-parametric approach to relax this assumption to an extent and maintain the computational advantages of BSL without any additional tuning.  We test our new method, semiBSL, on several challenging examples involving simulated and real data and demonstrate that semiBSL can be significantly more robust than BSL and another approach in the literature.
	\end{abstract}
\noindent
    {\it Keywords:} Likelihood-free inference, Approximate Bayesian computation (ABC), Copula, Nonparanormal distribution, Kernel density estimation, Robust estimation

    \newpage
	\section{Introduction} \label{sec:intro}
	
	Approximate Bayesian computation (ABC) is now a well-known method for conducting Bayesian statistical inference for stochastic simulation models that do not possess a computationally tractable likelihood function \citep{Sisson2018}.  ABC bypasses likelihood evaluations by preferring parameter configurations that generate simulated data close to the observed data, typically on the basis of a carefully chosen summarisation of the data.
	
	Although ABC has allowed practitioners in many diverse scientific fields to consider more realistic models, it still has several drawbacks.  ABC is effectively a nonparametric procedure, and scales poorly with summary statistic dimension.  Consequently, most ABC analyses resort to a low dimensional summary statistic to maintain a manageable level of computation, which could lead to significant information loss.  Secondly, ABC requires the user to select values for various tuning parameters, which can impact on the approximation.
	
	An alternative approach, called synthetic likelihood \citep[SL,][]{Wood2010,Price2018}, assumes that the model statistic for a given parameter value has a multivariate normal distribution.  The mean and covariance matrix of this distribution are estimated via independent simulations of the statistic, which is used to approximate the summary statistic likelihood at the observed statistic.  \citet{Price2018} developed the first Bayesian approach for SL, called Bayesian synthetic likelihood (BSL).  The parametric assumption made by the SL allows it to scale more efficiently to increasing dimension of the summary statistic \citep{Price2018}.  Further, the BSL approach requires very little tuning and is ideal for exploiting parallel computing architectures compared to ABC.
	
	The SL method has been tested successfully and shows great potential in a wide range of application areas such as epidemiology and ecology \citep{Barbu2017,Price2018}.  Further, \citet{Price2018} and \citet{Everitt2017} demonstrate that BSL exhibits some robustness to a departure from normality.  However, a barrier to the ubiquitous use of SL is its strong normality assumption.   We seek an approach that maintains the efficiency gains of BSL but enhances its robustness.
	
	The main aim of this paper is to develop a more robust approach to BSL.  We do this by developing a semi-parametric method for approximating the summary statistic likelihood, called semiBSL, which involves using kernel density estimates for the marginals and combining them with a Gaussian copula.  An incidental contribution of this paper is a thorough empirical investigation into the robustness of BSL.  We demonstrate through several simulated and real examples that semiBSL can offer significantly improved posterior approximations compared to BSL. Furthermore, we find that the number of model simulations required for semiBSL does not appear to increase significantly compared to BSL, and does not require any additional tuning parameters. However, we also find that the standard BSL approach can be remarkably robust on occasions.  We also identify the limits of semiBSL by considering an example with nonlinear dependence structures between summaries.  
	
	There have been other approaches developed for robustifying the synthetic likelihood. \citet{Fasiolo2018} developed an extended empirical saddlepoint (EES) estimation method, which involves shrinking the empirical saddlepoint approximation towards the multivariate normal distribution.  The regularisation helps ensure that their likelihood estimator is well-defined even when the observed statistic lies in the tail of the model summary statistic distribution. The shrinkage parameter (called the ``decay") needs to be selected by the user, and effectively represents a bias/variance trade-off.  More shrinkage towards the Gaussian decreases variance but increases bias. Although \citet{Fasiolo2018} consider frequentist estimation only, we demonstrate that it is straightforward to consider it within a Bayesian algorithm, allowing direct comparison with our new approach.  We demonstrate how our approach tends to outperform the saddlepoint approach in terms of accuracy, at least in our test examples we consider, with less tuning.
   
	Another alternative to synthetic likelihood is the approach of \citet{Dutta2017}, which frames the problem of estimating an intractable likelihood as ratio estimation.  After choosing an appropriate density for the denominator, the likelihood is estimated via supervised binary classification methods where pseudo datasets drawn from the likelihood are labelled with a ``1'' and pseudo datasets drawn from the denominator distribution are labelled with a ``0''.  The features of the classification problem are the chosen summary statistics and possibly transformations of them.  \citet{Dutta2017} use logistic regression for the classification task.  This approach has the potential to be more robust than the synthetic likelihood.  However, we do not compare with this approach for several reasons.  Firstly, \citet{Dutta2017} consider point estimation, and extension to Bayesian algorithms is not trivial.  Secondly, their method requires the practitioner to make several choices, making it difficult to perform a fair comparison.  Finally, their approach needs to perform a potentially expensive penalised logistic regression for each proposed parameter, making it more valuable for very expensive model simulators.

	\citet{Li2017} also use copula modelling within ABC.  However, they use a Gaussian copula for approximating the ABC posterior for a high-dimensional parameter, whereas we use it for modelling a high-dimensional summary statistic.
	
	The article is structured as follows.  Section \ref{sec:BSL} introduces relevant background and previous research on BSL. Section \ref{sec:semiBSL} presents our new method, semiBSL, for robustifying BSL. Section \ref{sec:app} shows the comparison of the performance of BSL and semiBSL with four simulated examples and one real data example with models of varying complexity and dimension.  Section \ref{sec:conc} concludes the paper, and points out the limitations of our approach and directions for future work.

	\section{Bayesian Synthetic Likelihood} \label{sec:BSL}
	
	In ABC and BSL, the objective is to simulate from the summary statistic posterior given by
	\begin{align*}
		p(\vect{\theta}|\vect{s_y}) \propto p(\vect{s_y}|\vect{\theta})p(\vect{\theta}),
	\end{align*}
	where $\vect{\theta} \in \Theta \subseteq \mathbb{R}^p$ is the parameter that requires estimation with corresponding prior distribution $p(\vect{\theta})$.  Here, $\vect{y} \in \mathcal{Y}$ is the observed data that are subsequently reduced to a summary statistic $\vect{s_y} = S(\vect{y})$ where $S(\cdot): \mathcal{Y} \rightarrow \mathbb{R}^d$ is the summary statistic function.  The dimension of the statistic $d$ must be at least the same size as the parameter dimension, i.e.\ $d \geq p$.
	
	The SL \citep{Wood2010} involves approximating $p(\vect{s_y}|\vect{\theta})$ with
	\begin{align}
		p(\vect{s_y}|\vect{\theta}) \approx p_A(\vect{s_y}|\vect{\theta}) = \mathcal{N}(\vect{s_{y}} | \vect{\mu}(\vect{\theta}),\vect{\Sigma}(\vect{\theta})). \label{eq:sl}
	\end{align}
	The mean and covariance $\vect{\mu}(\vect{\theta})$ and $\vect{\Sigma}(\vect{\theta})$ are not available in closed form but can be estimated via independent model simulations at $\vect{\theta}$.  The procedure involves drawing $\vect{x}_{1:n} = (\vect{x}_1,\ldots,\vect{x}_n)$, where $\vect{x}_i \stackrel{\mathrm{iid}}{\sim} p(\cdot|\vect{\theta})$ for $i=1,\ldots,n$, and calculating the summary statistic for each dataset, $\vect{s}_{1:n} = (\vect{s}_1,\ldots,\vect{s}_n)$, where $\vect{s}_i$ is the summary statistic for $\vect{x}_i,\ i=1,\ldots,n$.  These simulations can be used to estimate $\vect{\mu}$ and $\vect{\Sigma}$ unbiasedly 
	\small
	\begin{align}
		\begin{split}
			\vect{\mu}_n(\vect{\theta}) & = \frac{1}{n}\sum_{i=1}^{n}\vect{s}_{i}, \\  \vect{\Sigma}_n(\vect{\theta}) &= \frac{1}{n-1}\sum_{i=1}^{n}(\vect{s}_{i}-\vect{\mu}_n(\vect{\theta}))(\vect{s}_{i}-\vect{\mu}_n(\vect{\theta}))^{\top}.
			\label{eq:SLparams}
		\end{split}
	\end{align}
	\normalsize
	The estimates in \eqref{eq:SLparams} can be substituted into the SL in \eqref{eq:sl} to estimate the SL as $\mathcal{N}(\vect{s_{y}} | \vect{\mu}_n(\vect{\theta}),\vect{\Sigma}_n(\vect{\theta}))$.  We can sample from the approximate posterior using MCMC, see Algorithm \ref{alg:MCMCBSL}.  Theoretically, the corresponding MCMC algorithm targets the approximate posterior  $p_{A,n}(\vect{\theta}|\vect{s_y}) \propto \mathsf{E}[\mathcal{N}(\vect{s_{y}} | \vect{\mu}_n(\vect{\theta}),\vect{\Sigma}_n(\vect{\theta}))]p(\vect{\theta})$.    For finite $n$, $\mathcal{N}(\vect{s_{y}} | \vect{\mu}_n(\vect{\theta}),\vect{\Sigma}_n(\vect{\theta}))$ is a biased estimate of $\mathcal{N}(\vect{s_{y}} | \vect{\mu}(\vect{\theta}),\vect{\Sigma}(\vect{\theta}))$, thus resulting in $p_{A,n}(\vect{\theta}|\vect{s_y})$ being theoretically dependent on $n$ \citep{Andrieu2009}.   However, \citet{Price2018} demonstrate empirically that the BSL posterior $p_{A,n}(\vect{\theta}|\vect{s_y})$ is remarkably insensitive to its only tuning parameter, $n$.  Thus we can choose $n$ to maximise computational efficiency. 
	
	\begin{algorithm}[htp]
		\SetKwInOut{Input}{Input}
		\SetKwInOut{Output}{Output}
		\Input{Summary statistic of the data, $\vect{s_y}$, the prior distribution, $p(\vect{\theta})$, the proposal distribution $q$, the number of iterations, $T$, and the initial value of the chain $\vect{\theta}^{0}$.}
		\Output{MCMC sample $(\vect{\theta}^{0},\vect{\theta}^{1}, \ldots, \vect{\theta}^{T})$ from the BSL posterior, $p_{A,n}(\vect{\theta}|\vect{s_{y}})$.  Some samples can be discarded as burn-in if required. }
		\vspace{0.5cm}
		Simulate $\vect{x}_{1:n} \stackrel{\mathrm{iid}}{\sim} p(\cdot|\vect{\theta}^{0})$ and compute $\vect{s}_{1:n}$\\
		Compute $\vect{\phi}^{0}=(\vect{\mu}_n(\vect{\theta}^{0}),\vect{\Sigma}_n(\vect{\theta}^{0}))$ using \eqref{eq:SLparams} \\
		\For{$i = 1$ to $T$}{
			Draw $\vect{\theta}^{*} \sim q(\cdot|\vect{\theta}^{i-1})$\\
			Simulate $\vect{x}_{1:n}^{*} \stackrel{\mathrm{iid}}{\sim} p(\cdot|\vect{\theta}^{*})$ and compute $\vect{s}_{1:n}^{*}$\\
			Compute $\vect{\phi}^{*}=(\vect{\mu}_n(\vect{\theta}^{*}),\vect{\Sigma}_n(\vect{\theta}^{*}))$ using \eqref{eq:SLparams} \\
			Compute \tiny$r=\min\left(1,\frac{\mathcal{N}(\vect{s_{y}} | \vect{\mu}_n(\vect{\theta}^*),\vect{\Sigma}_n(\vect{\theta}^*))p(\vect{\theta}^{*})q(\vect{\theta}^{i-1}|\vect{\theta}^{*})}{\mathcal{N}(\vect{s_{y}} | \vect{\mu}_n(\vect{\theta}^{i-1}),\vect{\Sigma}_n(\vect{\theta}^{i-1}))p(\vect{\theta}^{i-1})q(\vect{\theta}^{*}|\vect{\theta}^{i-1})}\right)$\normalsize\\
			\eIf{$\mathcal{U}(0,1)<r$}{
				Set $\vect{\theta}^{i}=\vect{\theta}^{*}$ and $\vect{\phi}^{i}=\vect{\phi}^{*}$\\
			}{
				Set $\vect{\theta}^{i}=\vect{\theta}^{i-1}$ and $\vect{\phi}^{i}=\vect{\phi}^{i-1}$\\
			}
		}
		\vspace{0.5cm}
		\caption{MCMC BSL algorithm.}
		\label{alg:MCMCBSL}
	\end{algorithm}

	The computational efficiency of BSL over ABC stems from the normality assumption. This approximation can work well in many applications. However, there are also several reasons to suggest why it may not be appropriate.  For a given parameter value, there may be model summary statistics with non-normal marginal distributions and nonlinear dependencies between marginals.  Further, the normality assumption may not be appropriate over all the non-negligible posterior mass.  Finally, as the dimension of the summary statistic grows it is likely that the multivariate normal assumption becomes less reasonable.  
	
	In the next section, we develop a method to increase the robustness of BSL, while largely retaining its computationally convenient properties.

	\section{Semi-Parametric Approach to Bayesian Synthetic Likelihood} \label{sec:semiBSL}
	
	Here we propose to use copula models to approximate the joint distribution of the model summary statistic.  The main appeal of copula models is the ability to model marginal distributions and joint dependence structures separately.  In this paper we use kernel density estimates \citep[KDEs,][]{Rosenblatt1956,Parzen1962} for modelling each marginal distribution.   KDEs are useful as they can be flexible and are nonparametric, eliminating the need for the user to select specific parametric forms for the marginals.  However, we do note that other univariate density estimators could be used in our approach.  After transforming the marginals based on the fitted KDEs, we attempt to capture the dependence structure between summaries by using a Gaussian copula.  The appeal of the Gaussian copula is its tractability, which makes it ideal for its repeated use within a Bayesian algorithm.  The density estimator we use is similar to the approach of \citet{Liu2009}.
	
	Of course, the standard SL approach is a special case of ours when the marginal distributions are assumed normal.  Given the semi-parametric nature of the density estimator we use, we refer to our approach as semiBSL.  We now describe the technical details of our method.

	\paragraph{Kernel density estimation}
	
	We model each univariate marginal distribution of the joint summary statistic using a KDE.  A KDE of an unknown distribution $f_X(x)$ for a continuous random variable $X \in \mathcal{X}$ can be defined based on $n$ independent and identically drawn samples $x_1,\ldots,x_n \in \mathcal{X}$ as
	
	\begin{equation*}
		\hat{f}_X(x) = \dfrac{1}{n} \sum_{i=1}^{n} K_h(x-x_i),
	\end{equation*}
	where $K_h(\cdot)$ is the kernel function, which has nonnegative support over its domain and integrates to $1$, i.e.\ $\int_{\mathcal{X}} K_h(x)dx = 1$.  There are a large variety of kernels to choose from in KDE, and here we present results using a Gaussian kernel because it is nonzero everywhere. Note that we also tested the Epanechnikov kernel \citep{Epanechnikov1969}, which is optimal in terms of asymptotic mean integrated squared error. The posterior distributions produced by both kernels are visually very similar in several of our test examples.
	
	We generally assume that each summary statistic is continuous and unbounded.  Large discrete summary statistics (e.g.\ counts) can be treated as continuous.  Various transformations can be used if summary statistics are bounded (e.g.\ log transform for statistics taking positive values only).

	The smoothness of the KDE is controlled by the bandwidth parameter $h$.  Unlike standard ABC which pre-specifies the bandwidth parameter to use in approximating the joint summary statistic likelihood, we use the bandwidth suggested by \citet{silverman2018}, i.e. $h = 0.9n^{-0.2} \min(\sigma, \text{interquantile} \text{range}/1.34)$, where $\sigma$ is the standard deviation of the distribution and can be estimated empirically. This is also the adopted implementation in the $\mathsf{density}$ function in the statistical software R. We use the same bandwidth for cumulative distribution function (CDF) estimation later but it is possible to use a different bandwidth for the CDF if desired.
	
	\paragraph{Copula}
	
	Suppose a random vector $\vect{X} = (X_1,\ldots,X_d)^{\top}$ has continuous  marginals. Sklar's theorem \citep{Sklar1959} states that the multivariate CDF of $\vect{X}$ can be uniquely defined by its marginal CDFs, denoted, $u_j=F_j(x_j),j=1,\ldots,d$, and the dependence structure is captured with a multivariate copula function, where $u_j$ is uniformly distributed.  A convenient choice of copula in many applications is the Gaussian copula, whose probability density function (PDF) can be written as
	
	\begin{equation*}
		c_{\vect{R}}(\vect{u}) = \dfrac{1}{\sqrt{|\vect{R}|}} \exp\Big\{-\dfrac{1}{2} \vect{\eta}^{\top}(\vect{R}^{-1}-\vect{I}_d)\vect{\eta}\Big\},
	\end{equation*}
	where $\vect{I}_d$ is a $d$-dimensional identity matrix, and $\eta_j = \Phi^{-1}(u_j),j=1,\ldots,d$, in which $\Phi^{-1}(\cdot)$ is the inverse CDF of the standard normal distribution. The sole parameter $\vect{R}$ of the Gaussian copula is a correlation matrix, estimation of which will be discussed shortly. Once an estimated correlation matrix $\hat{\vect{R}}$ is obtained, semiBSL estimates the summary statistic likelihood $p(\vect{s}_{\vect{y}}|\vect{\theta})$ by
	
%	\scriptsize
	\begin{equation}
		g(\vect{s}_{\vect{y}}|\vect{\theta}) = \dfrac{1}{\sqrt{|\vect{\hat{R}}|}} \exp\Big\{-\dfrac{1}{2} \hat{\vect{\eta}}^{\top}_{\vect{s}_{\vect{y}}}(\vect{\hat{R}}^{-1}-\vect{I}_d)\hat{\vect{\eta}}_{\vect{s}_{\vect{y}}}\Big\} \prod_{j=1}^{d} \hat{f}_j(s_{y}^{j}), \nonumber\\
		\label{eq:pdf_semiBSL}
	\end{equation}
%	\normalsize
	where $s_{y}^{j}$ is the $j$th component of $\vect{s}_{\vect{y}}$, $\hat{f}_j(\cdot)$ is the estimated KDE of the $j$th marginal probability density,  $\hat{\vect{\eta}}_{\vect{s}_{\vect{y}}} = (\hat{\eta}_{s_{y}^{1}},\ldots,\hat{\eta}_{s_{y}^{d}})^\top$,
	$\hat{\eta}_{s_{y}^{j}} = \Phi^{-1}(\hat{u}_{j})$,  $j=1,\ldots,d$ and $\hat{u}_{j} = \hat{F}_j(s_{y}^{j})$.
	
	It is critical that the semi-parametric density estimator is fast to fit here, since it must be applied at every iteration of semiBSL.  The major appeal of the Gaussian copula here is its estimation tractability.  We discuss some other possible extension to the copula model in Section \ref{sec:conc}.

	\paragraph{Gaussian rank correlation}
	
	The standard approach to estimate the Gaussian copula parameter $\vect{R}$ is to compute the sample correlation matrix based on the collection $\{\vect{\eta}_{\vect{s}_i}\}_{i=1}^n$ where $\vect{\eta}_{\vect{s}_i} = (\eta_{s_i^1},\ldots,\eta_{s_i^d})^\top$ and $\eta_{s_i^j} = \Phi^{-1}(\hat{F}_i(s_i^j))$.  However, this approach for estimating $\vect{R}$ relies on each of the KDEs providing a very good fit to the actual distribution of the corresponding marginal.  Although KDE models are flexible, large sample sizes are often required to capture strong irregularities, even in univariate distributions.
		
	To obtain a procedure that is more robust to a potential lack of fit of the KDEs, we consider a nonparametric estimator of $\vect{R}$ using the Gaussian rank correlation \citep[GRC,][]{Boudt2012}.  Using the notation defined earlier, the $(i,j)$th component of the GRC estimate is given by
	
	\begin{equation}
		\hat{\rho}_{i,j}^{\mathrm{grc}} = \dfrac{\sum_{k=1}^n \Phi^{-1}\left(\dfrac{r(s_k^i)}{n+1}\right) \Phi^{-1}\left(\dfrac{r(s_k^j)}{n+1}\right)}
		{\sum_{k=1}^n \Phi^{-1} \left( \dfrac{k}{n+1} \right)^2},
		\label{eq:grc}
	\end{equation}
	where $r(\cdot): \mathbb{R} \rightarrow \mathcal{A}$, where $\mathcal{A} \equiv \{1,\ldots,n\}$, is the rank function. The GRC is not expensive to compute and has two favourable properties. \citet{Boudt2012} show that the GRC is robust to a small number of outliers. In fact, it requires at least $12.4\%$ of data contamination to revert a positive linear correlation. In addition, the GRC estimator is always semi-positive definite when data are multivariate normal. 
	
	The full procedure to estimate the summary statistic likelihood at the point $\vect{s}_{\vect y}$ using our semiBSL approach is shown in Algorithm \ref{alg:pdf_semiBSL}. Importantly, our method does not require any additional tuning parameters compared with BSL. Replacing the Gaussian estimate $\mathcal{N}(\vect{s_{y}} | \vect{\mu}_n(\vect{\theta}^*),\vect{\Sigma}_n(\vect{\theta}^*))$ with the semi-parametric estimate $g(\vect{s}_{\vect{y}} | \vect{\theta})$ of equation \eqref{eq:pdf_semiBSL} in Algorithm \ref{alg:MCMCBSL} gives MCMC semiBSL.
	
	\begin{algorithm}[htp]
		\SetKwInOut{Input}{Input}
		\SetKwInOut{Output}{Output}
		\Input{Collection of simulated summary statistics based on parameter $\vect{\theta}$, $\{\vect{s}_i\}_{i=1}^n$, and the summary statistic of the observed data, $\vect{s_y}$.}
		\Output{semiBSL estimate $g(\vect{s}_{\vect y}|\vect{\theta})$}
		\vspace{0.5cm}
		\For{$j = 1$ to $d$}{
			Estimate $\hat{f}_j(\vect{s}_{\vect{y}_j})$ and $\hat{F}_j(\vect{s}_{\vect{y}_j})$ based on $\{s_i^j\}_{i=1}^{n}$ using KDE.\\
		}
		\For{$i=2$ to $d$}{
			\For{$j=1$ to $d-1$}{
				Compute ranks $r(s_{k}^{i})$ and $r(s_{k}^{j})$ for $k=1,\ldots,n$.\\
				Estimate $\hat{\rho}_{i,j}^{\mathrm{grc}}$ using the ranks above with equation \eqref{eq:grc}.\\
			}
		}
		Construct $\hat{\vect{R}}$ and compute $g(\vect{s}_{\vect y} | \vect{\theta})$ with equation \eqref{eq:pdf_semiBSL}.
		\vspace{0.5cm}
		\caption{The semi-parametric procedure to estimate the summary statistic likelihood $g(\vect{s_y}|\vect{\theta})$ in semiBSL.}
		\label{alg:pdf_semiBSL}
	\end{algorithm}

	 Similar to BSL, the number of model simulations $n$ used in semiBSL represents a trade-off between the cost per iteration and the overall acceptance rate in MCMC.  In Appendix \ref{app:computational_efficiency_example}, we show that semiBSL does not require a larger $n$ compared to BSL in an illustrative toy example.  This is important, since we find in this example and the applications below that semiBSL can provide robustness without incurring additional model simulations.

	\section{Applications} \label{sec:app}
	
	In this section, we investigate the performance of semiBSL on five examples. All the examples have some noticeable non-normal marginal summary statistic regardless of the joint distribution, see Appendix \ref{app:subsec:dist_summStat}. We also look into an example with strong nonlinear dependencies between summaries, which challenges the Gaussian copula assumption in semiBSL. In addition to the results presented in this section, estimates of univariate posterior distributions are provided in Appendix B.2.
	
    We compare semiBSL with BSL and also a Bayesian version of the EES approach of \citet{Fasiolo2018}.  The EES method requires specification of an additional tuning parameter (the ``decay" parameter), which mutates the EES between an empirical saddlepoint approximation (decay equals to $0$) and a multivariate normal distribution (decay approaches infinity). We use the cross validation method suggested by \citet{Fasiolo2018} to select the decay, which is also available in the associated R package.
    
	As shown in \citet{Price2018} and \citet{An2018}, BSL is remarkably insensitive to $n$ in a variety of applications. We tested several different values of $n$ for each example, and found that semiBSL appears to inherit this useful property (see Appendix \ref{app:sensitivity2n} for more results).  Therefore, the number of simulations per iteration $n$ in all examples is chosen to maximise computational efficiency (measured by scaled effective sample size). The total number of simulations (the number of simulations per iteration times the total number of iterations) is fixed for BSL and semiBSL in all following examples. However since the performance of EES is not as stable, we increase the total number of simulations accordingly.

	We use MCMC with a multivariate normal random walk proposal to sample from the approximate posterior distribution in all methods. Since the different approaches may result in different posterior approximations, we tune the covariance matrix of the random walk separately.  We use pilot runs until we believe we have a reasonable estimate of the corresponding approximate posterior covariance that we use in the main MCMC run. In some examples, we use transformations so that the random walk takes place in an unconstrained parameter space. 

    The semiBSL method in this paper is implemented with the R package $\mathsf{BSL}$, which is available on github \textit{github.com/ziwenan/BSL}.
	
	\subsection{MA(2)} \label{subsec:ma2}
	
	The moving average (MA) time-series model is a popular toy example in ABC-related research topics \citep[e.g.][]{Chiachio2014} for its simplicity of simulation and availability of the likelihood function. We consider an MA(2) model with the following evolution for the data
	
	\begin{equation*}
		y_t = z_t + \theta_1 z_{t-1} + \theta_2 z_{t-2},
	\end{equation*}
	for $t = 1,\ldots,50$, where $z_t \sim \mathcal{N}(0,1)$, $t=-1,\ldots,50$. Constraints for $\vect{\theta}$ are $-1<\theta_2<1,\theta_1+\theta_2>-1,\theta_1-\theta_2<1$. Here, the likelihood function is multivariate normal with $\mathrm{Var}(y_t)=1+\theta_1^2+\theta_2^2$, $\mathrm{Cov}(y_t,y_{t-1})=\theta_1+\theta_1\theta_2$, $\mathrm{Cov}(y_t,y_{t-2})=\theta_2$, with all other covariances equal to $0$. Here, the observed dataset is simulated with parameters $\theta_1 = 0.6$ and $\theta_2 = 0.2$.
	
	We do not perform any data summarisation here, i.e.\ the summary statistic is the full dataset. Given the likelihood is multivariate normal, standard BSL will perform well. To explore its performance when normality is violated, we apply the sinh-arcsinh transformation \citep{Jones2009} to each individual data point to test different amounts of skewness and kurtosis,
	
		\begin{equation}
			f(x) = \sinh \Big(\dfrac{1}{\delta} \sinh^{-1}(x + \epsilon) \Big),
			\label{eq:transformation}
		\end{equation}
	where $\epsilon$ and $\delta$ control the level of skewness and kurtosis, respectively. We tested the following four scenarios: no transformation, skewness only, kurtosis only and both skewness and kurtosis (see Figure \ref{fig:summStat_ma2} in Appendix \ref{app:computational_efficiency_example} for an example of the distributions). In the last experiment, we use different transformation parameters for each marginal summary statistic to aggravate estimation difficulty, i.e.\ $\epsilon_i \stackrel{\mathrm{iid}}{\sim} \mathrm{U}(-2,2)$, $\delta_i = u_i ^ {(-1)^{v_i}}$, where $u_i \stackrel{\mathrm{iid}}{\sim} \mathrm{U}(1,2)$, $v_i \stackrel{\mathrm{iid}}{\sim} \mathrm{Binary}(0.5)$. 
	
\begin{figure*}
		\centering
		\includegraphics[width=0.8\textwidth]{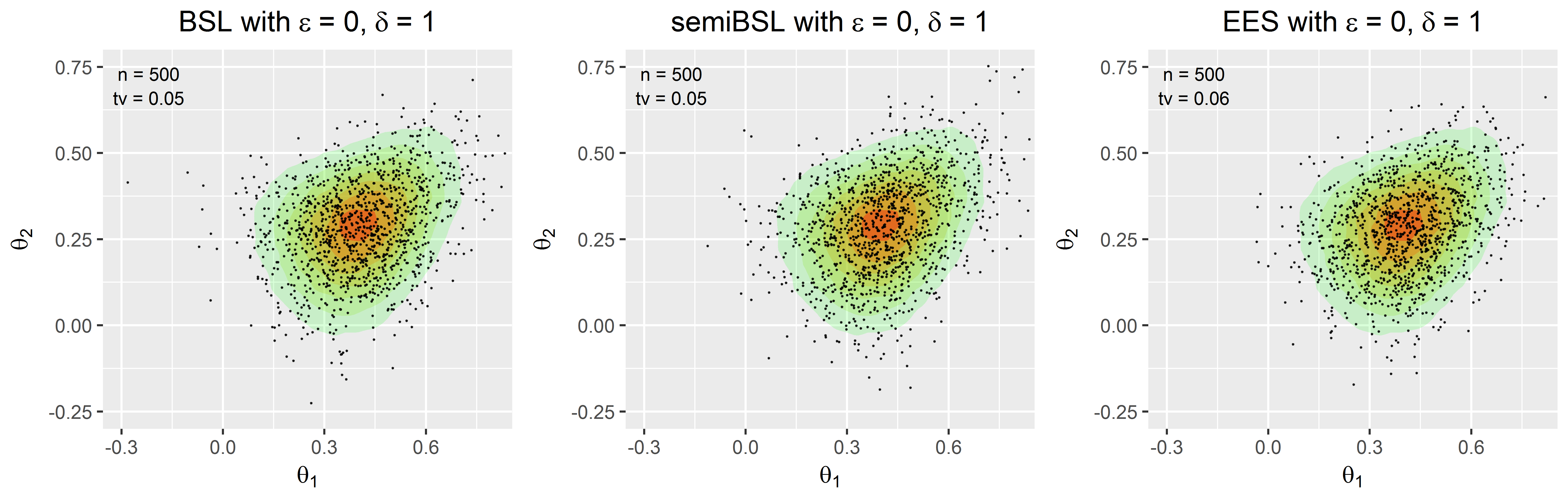}
		\includegraphics[width=0.8\textwidth]{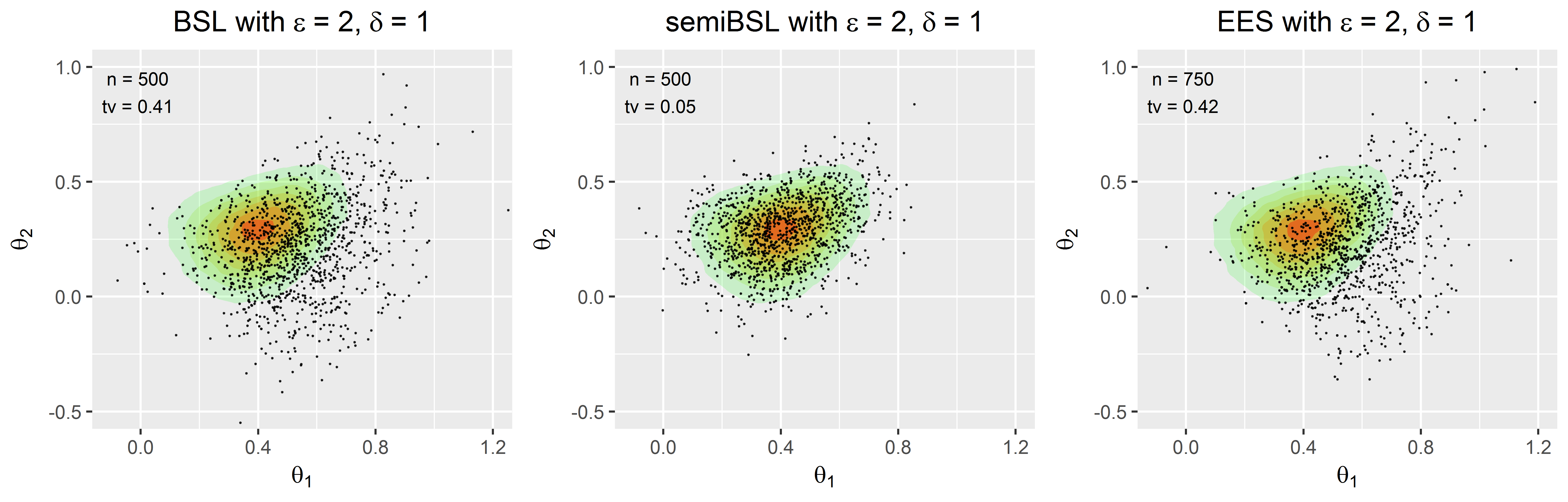}
		\includegraphics[width=0.8\textwidth]{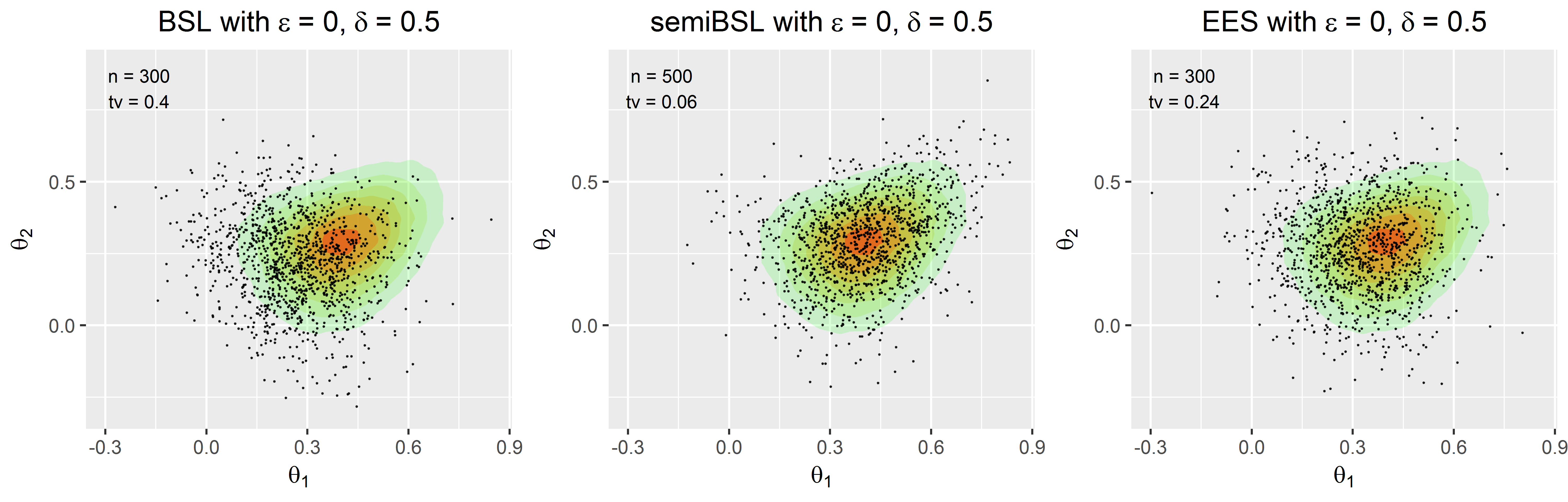}
		\includegraphics[width=0.8\textwidth]{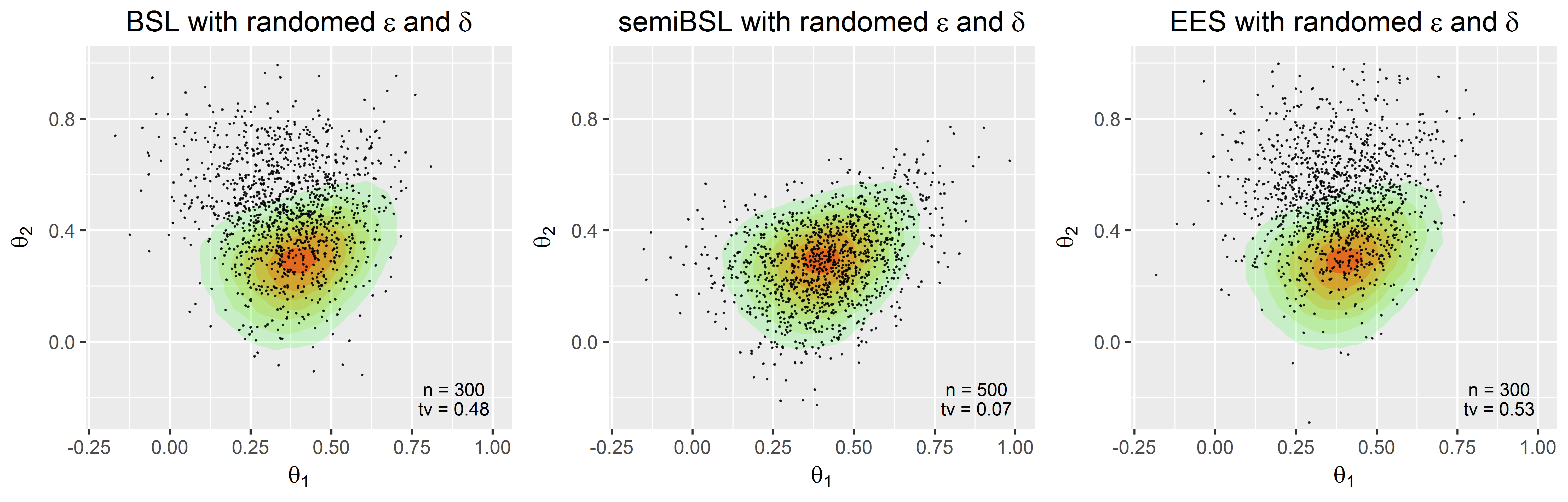}
		\caption{Bivariate scatter plots of the posterior distributions of the MA(2) example using different pairs of transformation parameters for the summary statistic. The overlaid contour plots represent the true posterior. The number of simulations per iteration $n$ and the total variation distance compared to the gold standard are shown in the corner of each plot.}
		\label{fig:scatter_ma2}
	\end{figure*}

	Estimated posterior densities are presented in Figure \ref{fig:scatter_ma2}.  The results are shown for four different transformations and three different posterior approximations (BSL, semiBSL and EES).   The contour plots represent the ``true" posterior (obtained from a long run of MCMC using the exact likelihood).  For the posterior approximations, bivariate scatterplots of the posterior samples are shown. We use the total variation distance to quantify the disparity between the approximated posterior distribution and the true posterior, i.e.\ $\text{tv}(f_1,f_2) = \dfrac{1}{2} \int_{\vect{\Theta}} |f_1(\vect{\theta}) - f_2(\vect{\theta})| d \vect{\theta}$. This can be estimated via numerical integration with a 2-D grid covering most of the posterior mass. The choice of $n$ and the total variation distance are shown in Figure \ref{fig:scatter_ma2}.
	
	In order to show the stability and performance of our approximate posterior results, we also run BSL and semiBSL for additional datasets generated from the MA(2) model with the same true parameter configuration. We compute the total variation using nonparametric density estimation for each case and summarise the results in Table \ref{tab:tv_ma2}. It is apparent that semiBSL always outperforms BSL in terms of the total variation distance to the true posterior.
	
	\begin{table} \label{tab:tv_ma2}
		\centering
		\caption{Total variation distance from the approximate posterior to the true posterior for the MA(2) example. The values shown in the table are the mean of the total variation distances of $30$ different replications. The standard errors are given in the parentheses.}
		\vspace{0.5cm}
		\begin{tabular}{lll}
			\hline\noalign{\smallskip}
			$\epsilon$ and $\delta$ & tv (BSL) & tv (semiBSL)  \\
			\noalign{\smallskip}\hline\noalign{\smallskip}
			$\epsilon = 0$, $\delta = 1$ & 0.07 (0.02) & 0.07 (0.02) \\
			$\epsilon = 2$, $\delta = 1$ & 0.41 (0.16) & 0.22 (0.09) \\
			$\epsilon = 0$, $\delta = 0.5$ & 0.42 (0.18) & 0.11 (0.03) \\
			random $\epsilon$ and $\delta$ & 0.47 (0.19) & 0.16 (0.08) \\
			\noalign{\smallskip}\hline
		\end{tabular}
	\end{table}
	
	The bivariate posteriors and total variations indicate that the semiBSL posteriors are robust to all four transformed summary statistics, while BSL fails to get close to the true posterior.  The results also suggest that the EES provides little robustness in this example. We also tested the EES method with a much larger $n$, here $n=5000$, for the randomised $\epsilon, \delta$ dataset to see whether the posterior accuracy improves.  With $n=5000$, the decay parameter is reduced to 45 (roughly one third of the value obtained with $n=300$).  We find that even with significant additional computational effort, the EES posterior approximation shows little improvement.
	
	It is important to note that the dependence structure in the data (after back-transformation) is Gaussian. Therefore, the Copula assumption made by semiBSL is correct in this example up to estimation of the marginals, which are done using KDE without knowledge of true marginals. Thus, we would expect semiBSL to provide good posterior approximations here and it is necessary to test its performance in other examples. However, the example does serve to illustrate that BSL can be impacted by non-normality and that the EES may not provide sufficient robustness to non-normality.

	\subsection{M/G/1} \label{subsec:mg1}
	
	The M/G/1 (single-server queuing system with Poisson arrivals and general service times) queuing model has been investigated previously in the context of ABC by \citet{Blum2010} and \citet{Fearnhead2012}. Here, the observed data, $\vect{y}_{1:50}$, are $50$ inter-departure times from $51$ customers. In this model, the likelihood is cumbersome to calculate whereas simulation is trivial. The distribution of the service time is uniform $\mathrm{U}(\theta_1,\theta_2)$, and the distribution of the inter-arrival time is exponential with rate parameter $\theta_3$. We take $\vect{\theta} = (\theta_1,\theta_2,\theta_3)$ as the parameter of interest and put a uniform prior $\mathrm{U}(0,10) \times \mathrm{U}(0,10) \times \mathrm{U}(0,0.5)$ on $(\theta_1,\theta_2-\theta_1,\theta_3)$. The observed data are generated from the model with true parameter $\vect{\theta} = (1,5,0.2)$.
	
\begin{figure*}
		\centering
		\includegraphics[width=0.8\textwidth]{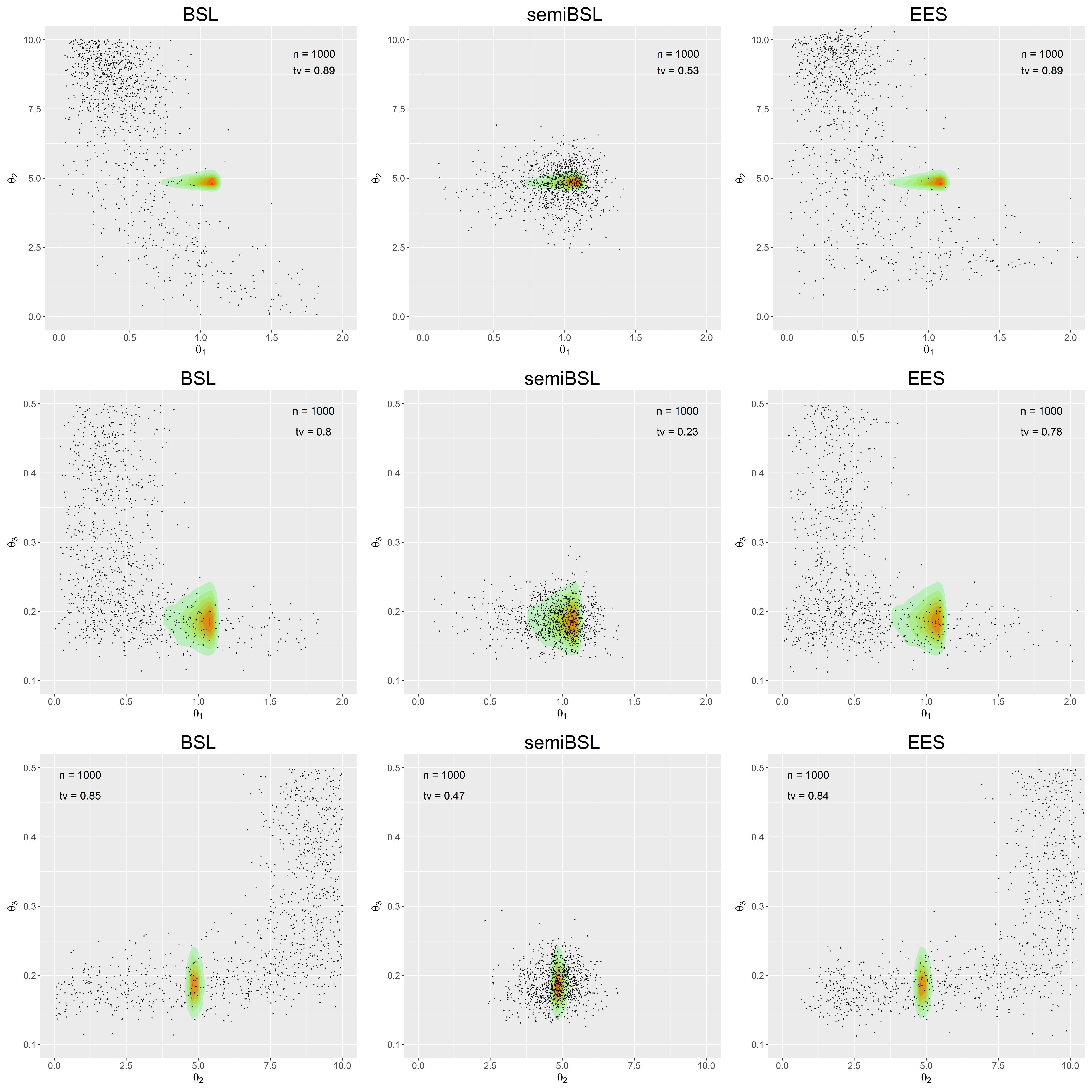}
		\caption{Bivariate scatter and contour plots of the posterior distributions of the M/G/1 example. The scatter plot is generated with a thinned approximate sample obtained by different approaches. The contour plot is drawn based on the ``true'' posterior distribution. The number of simulations per iteration $n$ and the total variation distance compared to the gold standard are shown in the corner of each plot.}
		\label{fig:scatter_mg1}
	\end{figure*}

	Here, we take the log of the inter-departure times as our summary statistic. It is interesting that the distribution of the summary statistics does not resemble any common distribution, see Figure \ref{fig:summStat_mg1} of Appendix \ref{app:subsec:dist_summStat}. With the Bayesian approach given by \citet{Shestopaloff2014}, we obtained a ``true'' posterior distribution of the model (sampled with MCMC). Figure \ref{fig:scatter_mg1} shows the bivariate scatterplot of posterior samples using BSL, semiBSL and EES. Here the number of simulations per iteration is $n=1000$ for all methods. The coloured contour plot indicates the true posterior. It is evident that both BSL and EES exhibit an ``L'' shape in the bivariate posterior and can hardly reflect the true posterior. The semiBSL posterior is significantly more accurate and provides reasonable estimates of the posterior means.
	
	Given that the posterior distribution of the M/G/1 example can be sensitive to the observed dataset, we repeat the MCMC BSL and MCMC semiBSL methods for $50$ different observed datasets using the same true parameter configuration and computed the 2D total variation distance to each true posterior for each pair of parameters. From Table \ref{tab:tv_mg1}, it is evident that semiBSL outperforms BSL in terms of posterior accuracy.
	
	\begin{table} \label{tab:tv_mg1}
		\centering
		\caption{Total variation distance from the approximate posterior to the true posterior for the M/G/1 example. The values shown in the table are the mean of the total variation distances of $50$ different replications. The standard errors are given in the parentheses.}
		\vspace{0.5cm}
		\begin{tabular}{lll}
			\hline\noalign{\smallskip}
			pair of parameters & tv (BSL) & tv (semiBSL)  \\
			\noalign{\smallskip}\hline\noalign{\smallskip}
			$\theta_1$, $\theta_2$ & 0.89 (0.07) & 0.55 (0.12) \\
			$\theta_1$, $\theta_3$ & 0.72 (0.05) & 0.38 (0.12) \\
			$\theta_2$, $\theta_3$ & 0.80 (0.10) & 0.42 (0.12) \\
			\noalign{\smallskip}\hline
		\end{tabular}
	\end{table}
    
	\subsection{Stereological Extremes} \label{subsec:stereo}
	
	During the process of steel production, the occurrence of microscopic particles, called inclusions, is a critical measure of the quality of steel. It is desirable that the inclusions are kept under a certain threshold, since steel fatigue is believed to start from the largest inclusion within the block. Direct observation of $3$-dimensional inclusions is inaccessible, so that inclusion analysis typically relies on examining $2$-dimensional planar slices. \citet{Anderson2002} establish a mathematical model to formulate the relationship between observed cross-section samples, $S$, and the real diameter of inclusions, $V$, assuming that the inclusions are spherical. The model focuses on large inclusions, i.e.\ $V > \nu_0$, where $\nu_0$ is a certain threshold, which is endowed with a generalised Pareto distribution such that
	
	\begin{equation*}
		P(V \leq \nu | V > \nu_0) = 1 - \left\{1 + \dfrac{\xi (\nu - \nu_0)}{\sigma}\right\} ^{-1/\xi} _{+},
	\end{equation*}
	where $\sigma > 0$ and $\xi \in \mathbb{R}$ are the scale and shape parameters, respectively, and $\{\cdot\}_{+} = \max\{0,\cdot\}$. The inclusions are mutually independent and locations of them follow a homogeneous Poisson process with rate parameter $\lambda$. While the spherical model possesses a likelihood function that is easily computable, the spherical assumption itself might be inappropriate. This leads to the ellipsoidal model proposed by \citet{Bortot2007}, who use ABC for likelihood-free inference due to the intractable likelihood function that it inherits. The new model assumes that inclusions are ellipsoidal with principal diameters $(V_1,V_2,V_3)$, where $V_3$ is the largest diameter without loss of generality. Here, $V_1 = U_1V_3$ and $V_2 = U_2V_3$, where $U_1$ and $U_2$ are independent uniform $\mathrm{U}(0,1)$ random variables. The observed value $S$ is the largest principal diameter of an ellipse in the two-dimensional cross-section.
	
	Here we consider the ellipsoidal model with parameter of interest $\vect{\theta} = (\lambda,\sigma,\xi)$. The prior distribution is $\mathrm{U}(30,200) \times \mathrm{U}(0,15) \times \mathrm{U}(-3,3)$. Denoting the observed samples as $\vect{S}$, we consider four summary statistics: the number of inclusions, $\log(\min(\vect{S}))$, $\log(\text{mean}(\vect{S}))$, $\log(\max(\vect{S}))$. Figure \ref{fig:summStat_stereo} in Appendix \ref{app:subsec:dist_summStat} shows the distribution of the chosen summary statistic simulated at a point estimate of $\vect{\theta}$. The last three summary statistics have a significantly heavy right tail, strongly invalidating the normality assumption of BSL.
	
\begin{figure}
		\centering
		\includegraphics[width=0.5\textwidth]{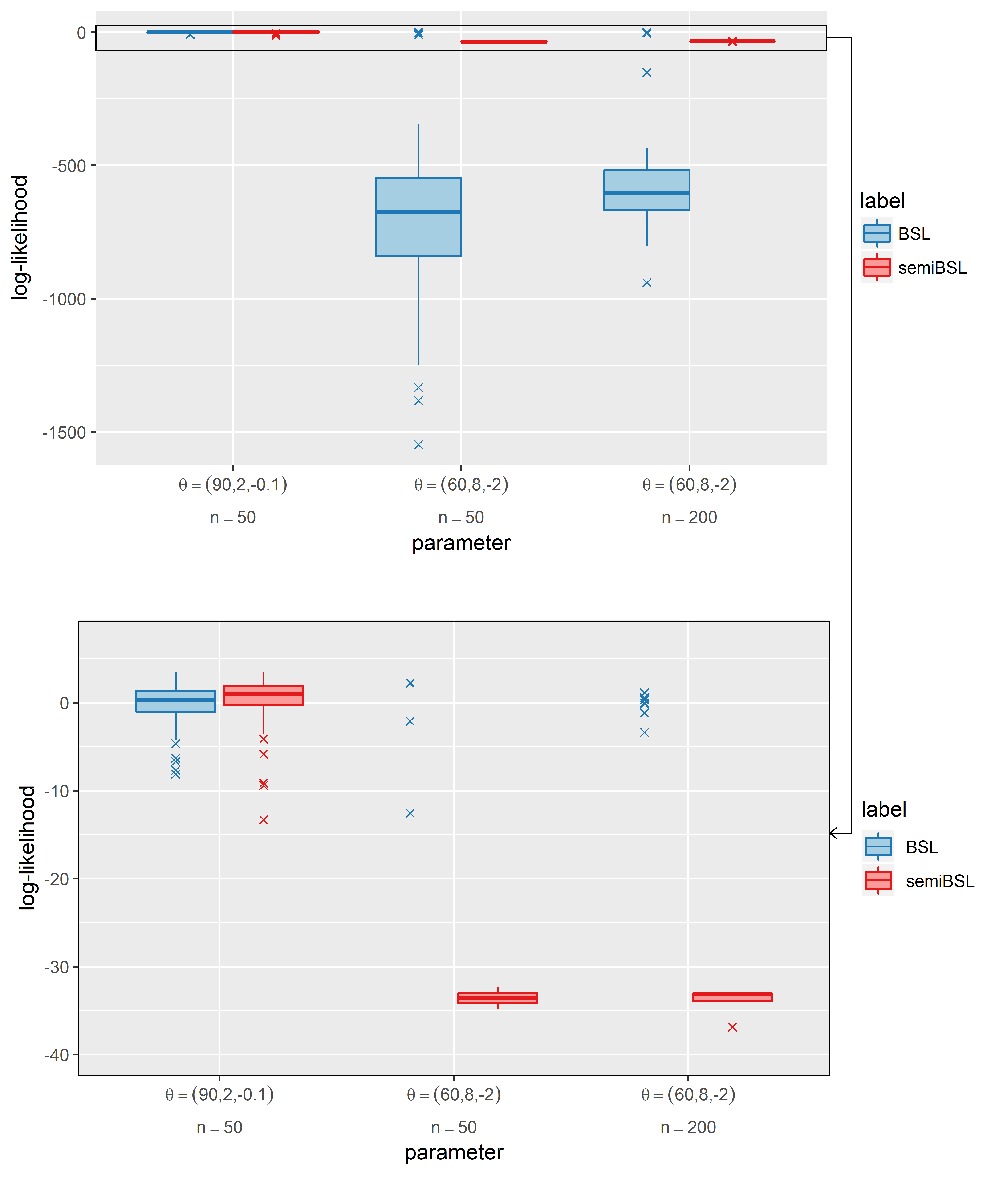}
		\caption{Boxplot of BSL and semiBSL log-likelihoods of the stereological extreme example. Negative infinite semiBSL log-likelihood values are ignored. The figure at top is the whole view of the boxplot. The figure at the bottom is zoomed in vertically to -40 to 7 to show more clearly highlight the outliers for BSL.}
		\label{fig:box_stereo}
	\end{figure}

\begin{figure*}
		\centering
		\includegraphics[width=0.8\textwidth]{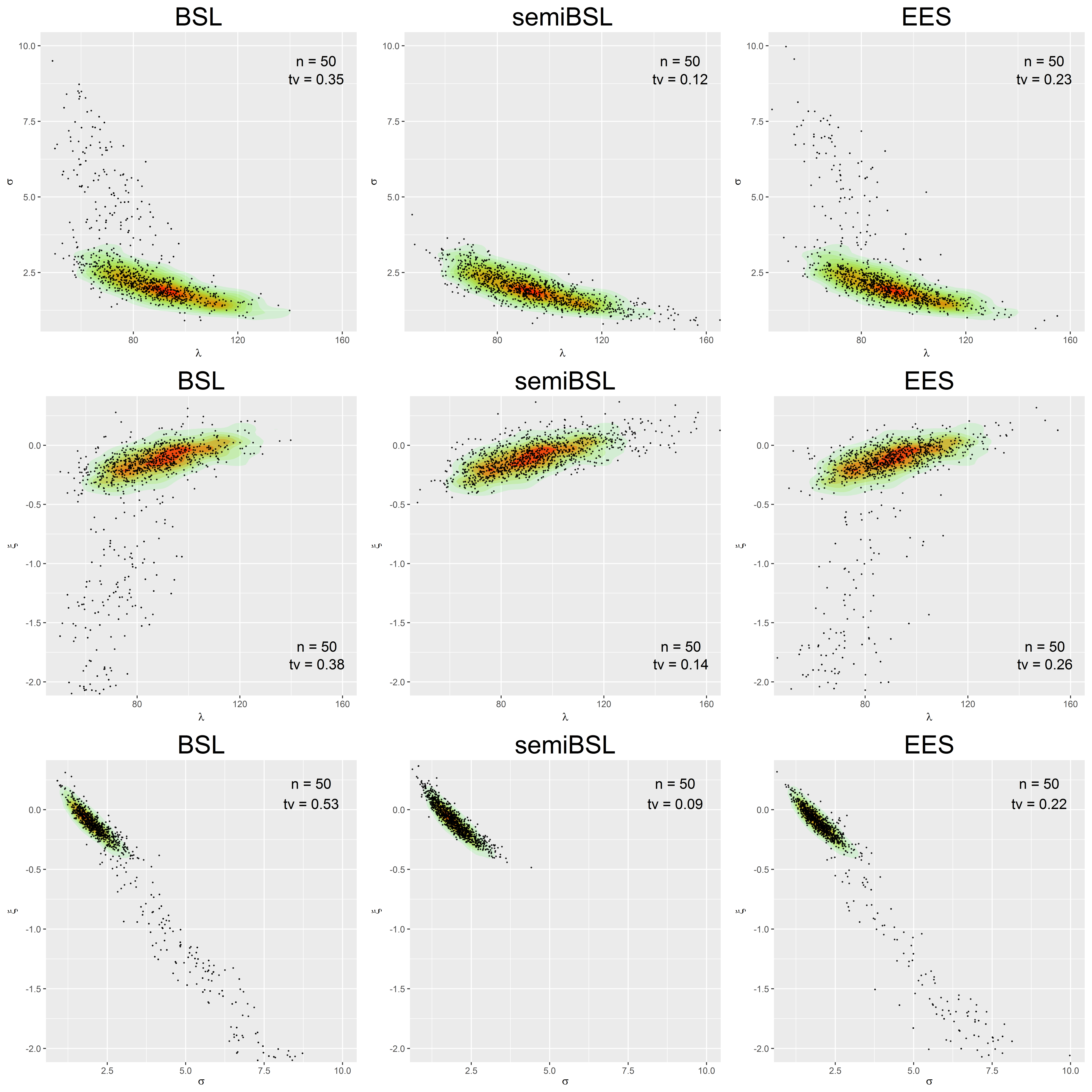}
		\caption{Bivariate scatter plots of the posterior distributions of the stereological extreme example. The scatter plot is generated with thinned approximate sample by BSL (run no.$2$ in Figure \ref{fig:post_stereo}), semiBSL and EES approach. The contour plot corresponds to the gold standard in this example (MCMC ABC). The number of simulations per iteration $n$ and the total variation distance compared to the gold standard are shown in the corner of each plot.}
		\label{fig:scatter_stereo}
	\end{figure*}

	Figure \ref{fig:scatter_stereo} shows the bivariate scatterplot of posteriors obtained by BSL, semiBSL and EES.  The number of simulations is $n=50$ for all methods. The overlaying contour plot is drawn with an MCMC ABC result with tolerance $1$. A Mahalanobis distance is used to compare summary statistics.  The covariance used in the Mahalanobis distance is taken to be the sample covariance of summary statistics independently generated from the model at a point estimate of the parameter. Note that the outliers are removed before computing the Mahalanobis covariance matrix. Given that there are only four summary statistics in this example, we take ABC as the gold standard approximation.  It is apparent that both BSL and EES results are accepting parameter values in the tails that are rejected by ABC and semiBSL. We use the boxplot (Figure \ref{fig:box_stereo}) to explain the impact of outlier simulations in BSL. The first column uses a parameter value that has high posterior support in the semiBSL result (referred to here as ``good") and a medium number of simulations, the second column uses a parameter value that should have negligible posterior support (referred to here as ``poor") and a medium number of simulations, and the last column uses a poor parameter value and a large number of simulations. In each test, we simulate $300$ independent BSL and semiBSL log-likelihood estimates. Negative infinities are ignored in the second and third tests in semiBSL. It is worth noting that semiBSL produces a lot of negative infinity log-likelihood estimates for simulations at the poor parameter value, which are ignored in the boxplot. The overestimated log-likelihoods at the ``poor'' parameter value by BSL are competitive to those at the ``good'' parameter value. Thus, when BSL grossly overestimates the likelihood at a poor parameter value, the algorithm may get stuck  in the tails for long periods. Overall, semiBSL is the only method that gives an approximate posterior consistent with that of ABC.
	
	\subsection{Fowler's Toads} \label{subsec:toads}
	
	Movements of amphibian animals exhibit patterns of site fidelity and long-distance dispersal at the same time. Modelling such patterns helps in understanding amphibian's travel behaviour and contributes to amphibian conservation. \citet{Marchand2017} develop an individual-based model for a species called Fowler's Toads (\textit{Anaxyrus fowleri}), and collect data via radiotracking in Ontario, Canada. The comprehensive experimental and modelling details are stated in the original paper. Here, we only present a brief description of the model.
	
	The model assumes that a toad hides in its refuge site in the daytime and moves to a randomly chosen foraging place at night. After its geographical position is collected via a transmitter, the toad either takes refuge at the current location or returns to one of the previous sites. For simplicity, the refuge locations are projected to a single axis, thus can be represented by a single-dimensional spatial process. GPS location data are collected on $n_t$ toads for $n_d$ days, i.e.\ the observation matrix $\vect{Y}$ is of dimension $n_d \times n_t$. For the synthetic data we use here, $n_t=66$, $n_d=63$, and missingness is not considered. Then $\vect{Y}$ is summarised down to four sets comprising the relative moving distances for time lags of $1,2,4,8$ days. For instance, $\vect{y}_1$ consists of the displacement information of lag $1$ day, $\vect{y}_1 = \{|\Delta y| = |\vect{Y}_{i,j}-\vect{Y}_{i+1,j}| ; 1 \leq i \leq n_d-1, 1 \leq j \leq n_t \}$.
	
	Simulation from the model involves two distinct processes. For each single toad, we first generate an overnight displacement, $\Delta y$, then mimic the returning behaviour with a simplified model. The overnight displacement is deemed to have significant heavy tails, assumed to belong to the L\'evy-alpha stable distribution family, with stability parameter $\alpha$ and scale parameter $\gamma$. This distribution has no closed form, while simulation from it is straightforward \citep{Chambers1976}, making simulation-based approaches appealing. The original paper provided three returning models with different complexity. We adopt the random return model here as it has the best performance among the three and is the easiest for simulation. The total returning probability is a constant $p_0$, and if a return occurs on day $1 \leq i \leq m$, then the return site is the same as the refuge site on day $i$, where $i$ is selected randomly from ${1,2,\ldots,m}$ with equal probability. Here we take the observed data as synthetically generated with true parameter $\vect{\theta}=(\alpha,\gamma,p_0)=(1.7,35,0.6)$. We use a uniform prior over $(1,2) \times (0,100) \times (0,0.9)$ here.
	
\begin{figure}
		\centering
		\includegraphics[width=0.75\textwidth]{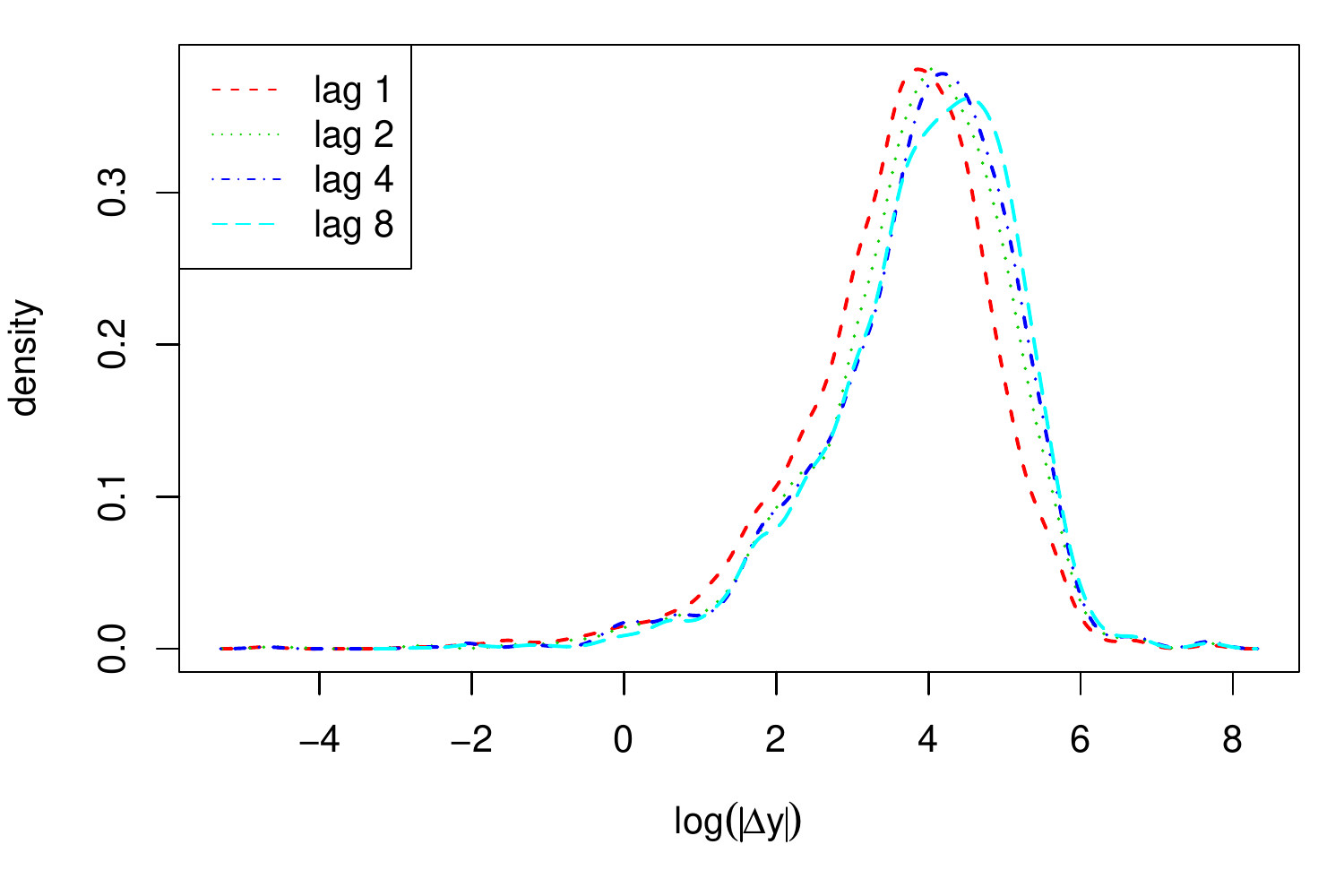}
		\caption{Density plot of $\log(|\Delta y|)$ for lags of $1,2,4,8$ days of the Fowler's toads example.}
		\label{fig:kde_deltay_toads}
	\end{figure}

	We fit a four-component Gaussian mixture model to each set of $\log(|\Delta y|)$. Figure \ref{fig:kde_deltay_toads} shows the distributions of $\log(|\Delta y|)$ for lags of $1,2,4,8$. As the summary statistic we use the 11-dimensional score of this fitted auxiliary model (corresponding to three component weights, four means and four standard deviations). This corresponds to the indirect inference approach for selecting summary statistics \citep[see][]{Drovandi2011,Gleim2013,Drovandi2015}. Accommodating the four different lags, there are $44$ summary statistics in total. The scores do not seem to depart a large amount from normality (see Figure \ref{fig:summStat_scores_toads} in Appendix \ref{app:subsec:dist_summStat}), thus standard BSL may be suitable for this model. To further explore the robustness of BSL and test our semiBSL approach, we include a power transformation of $\vect{s}_{\vect{x}}$ to push the irregularity of the summary statistic further. The transformation function is given by $f_{p}(\cdot) = \mathrm{sgn}(\cdot) \times (|\cdot|)^{p}$. It retains the sign of the input and creates a sharp peak near $0$. The distribution of the transformed summary statistics using $p=1.5$ is also included in Appendix \ref{app:subsec:dist_summStat}.
	
\begin{figure*}
		\centering
		\includegraphics[width=0.8\textwidth]{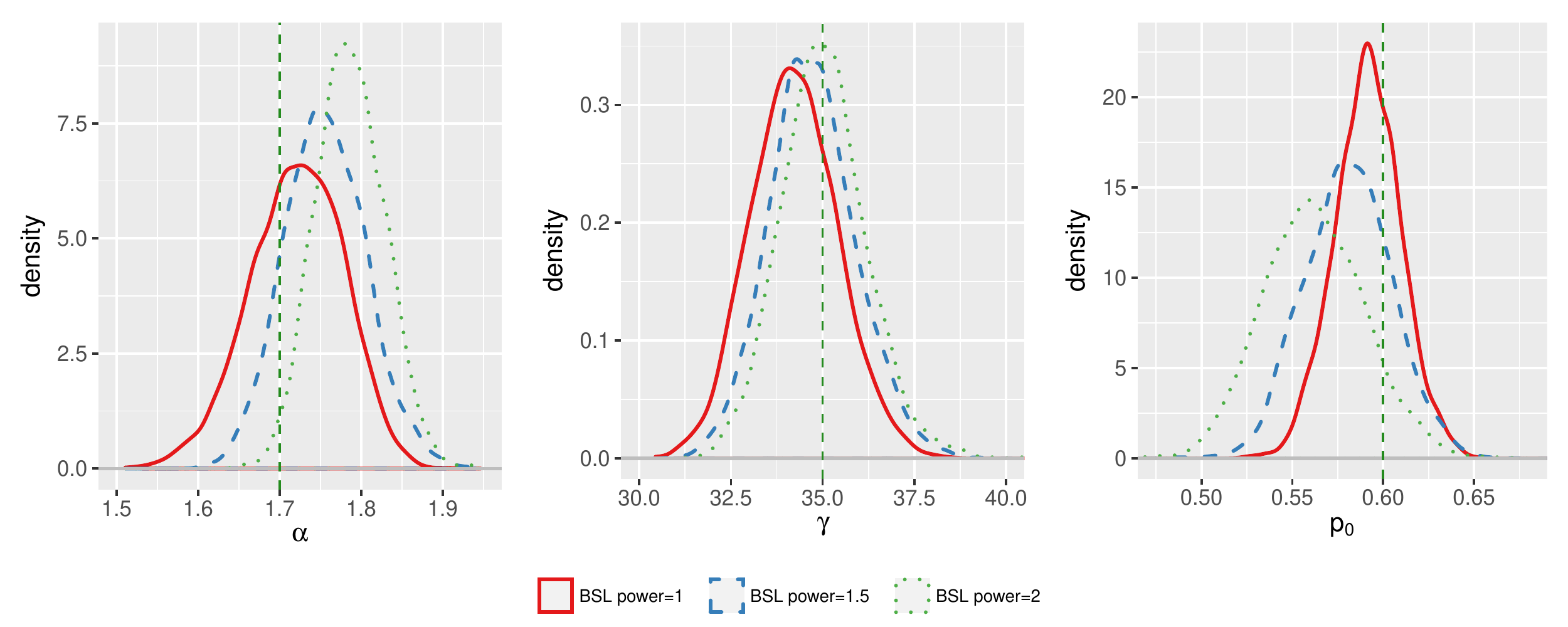}
		\includegraphics[width=0.8\textwidth]{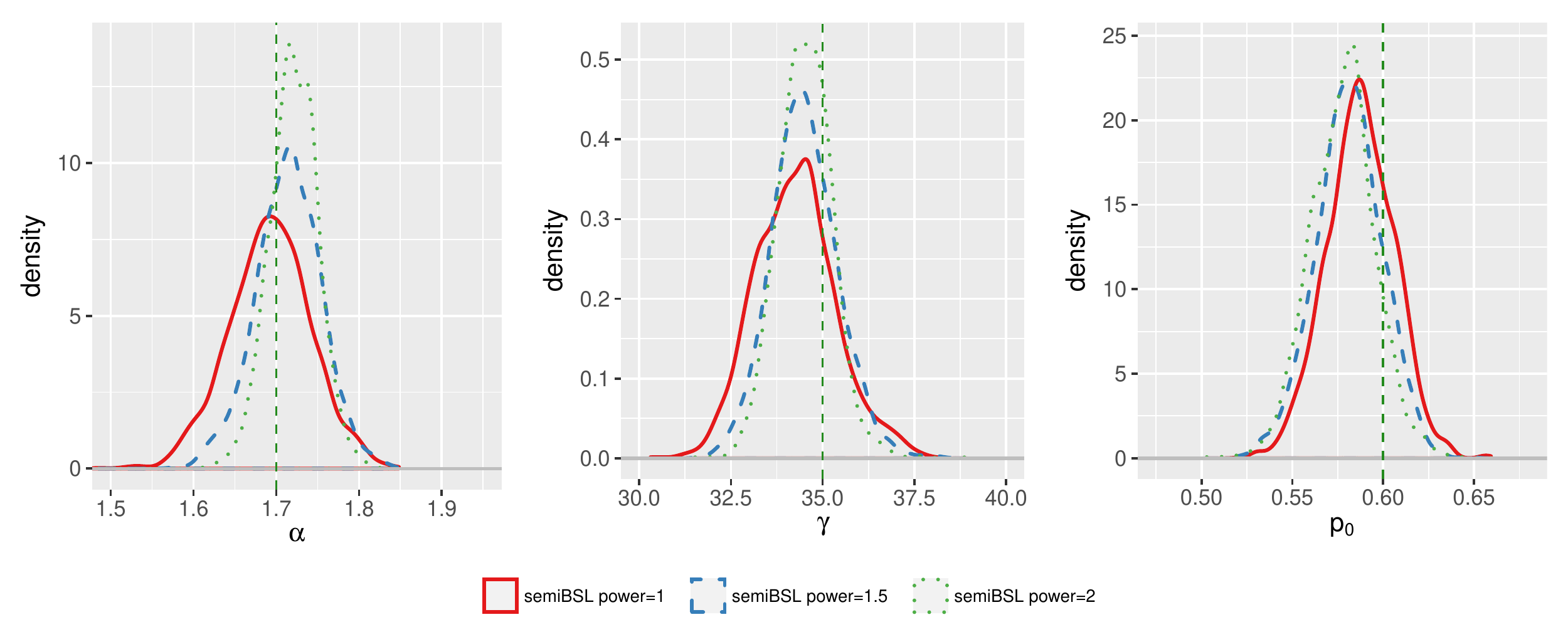}
		\includegraphics[width=0.8\textwidth]{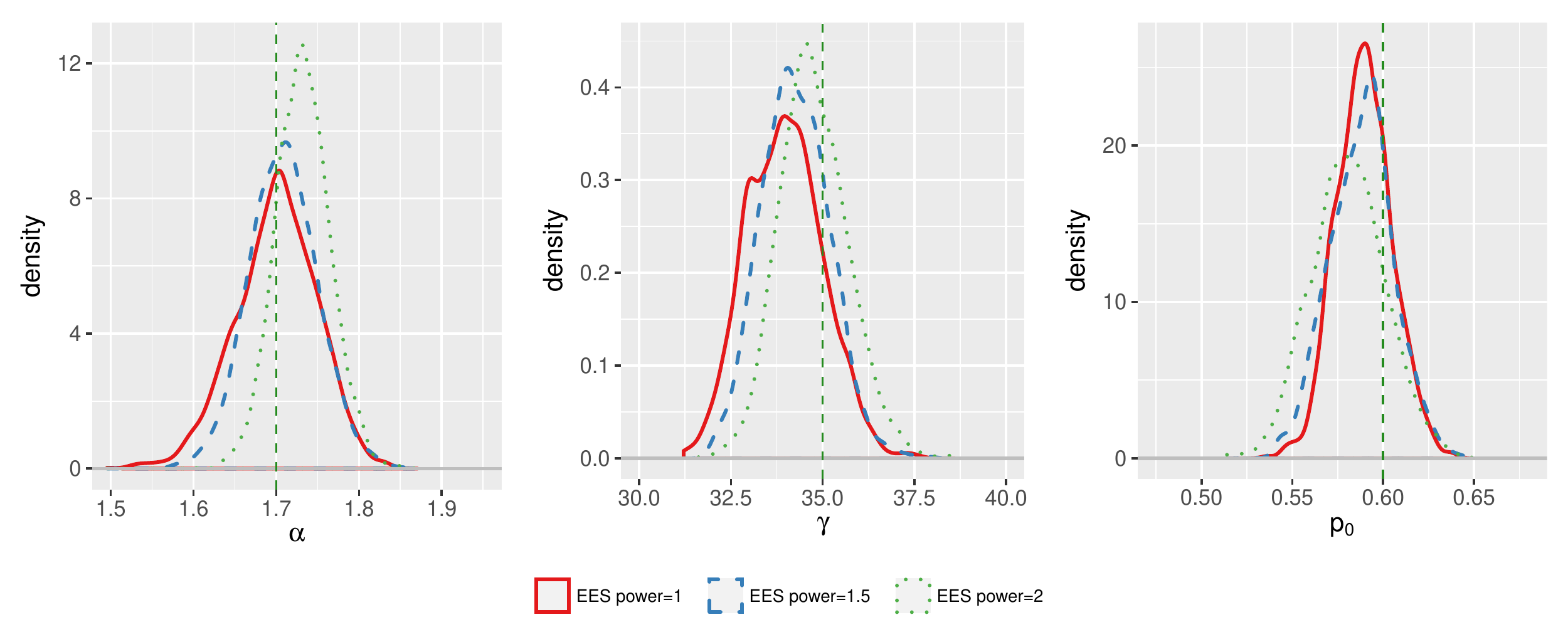}
		\caption{Comparing the approximate posterior distributions for the BSL, semiBSL and EES approaches at different transformation powers of the Fowler's toads example. The vertical line indicates the true parameter value.}
		\label{fig:post_comp_toads}
	\end{figure*}

	In Figure \ref{fig:post_comp_toads}, we show the posterior approximations produced by different approaches and transformation parameters. The posterior distributions obtained by different approaches are reasonably close to each other using the original score summary statistics. The BSL marginal posterior shows significant shift horizontally as the transformation power grows, whist semiBSL and EES are more robust to the change.
	
	\subsection{Simple Recruitment, Boom and Bust} \label{subsec:bnb}

	Here we consider an example that tests the limits of our semiBSL method. The simple recruitment, boom and bust model was used in \citet{Fasiolo2018} to investigate the performance of the saddlepoint approximation to a non-normal summary statistic. This is a discrete stochastic temporal model that can be used to represent the fluctuation of the population size of a certain group over time. Given the population size $N_t$ and parameter $\vect{\theta}=(r,\kappa,\alpha,\beta)$, the next value $N_{t+1}$ follows the following distribution
	
	\begin{align*}
		N_{t+1} \sim 
		\begin{cases}
			\mathrm{Poisson}(N_t(1+r)) + \epsilon_t, & \text{ if }\quad N_t \leq \kappa \\
			\mathrm{Binom}(N_t,\alpha) + \epsilon_t, & \text{ if}\quad N_t > \kappa
		\end{cases},
	\end{align*}
	where $\epsilon_t \sim \mathrm{Pois}(\beta)$ is a stochastic term. The population oscillates between high and low level population sizes for several cycles. The true parameters are $r=0.4$, $\kappa=50$, $\alpha=0.09$ and $\beta=0.05$, and the prior distribution is $\mathrm{U}(0,1) \times \mathrm{U}(10,80) \times \mathrm{U}(0,1) \times \mathrm{U}(0,1)$. There are $250$ values in the observed data. We use $50$ burn-in values to remove the transient phase of the process.
	
	We construct the summary statistics as follows. Consider a dataset $\vect{x}$, define the differences and ratios as $\vect{d}_{\vect{x}} = \{x_i - x_{i-1} ; i=2,\ldots,250\}$ and  $\vect{r}_{\vect{x}} = \{x_i / x_{i-1} ; i=2,\ldots,250\}$, respectively. We use the sample mean, variance, skewness and kurtosis of $\vect{x}$, $\vect{d}_{\vect{x}}$ and $\vect{r}_{\vect{x}}$ as our summary statistic, $\vect{s}_{\vect{x}}$. We also tested the statistics used in \citet{Fasiolo2018} but we found our choice to be more informative about the model parameters. The parameter $\beta$ seems to have a strong impact on the model statistic distribution. Small values of $\beta$ tend to generate statistics that are highly non-normal so we consider such a case here.
	
\begin{figure}
    	\centering
    	\includegraphics[width=0.75\textwidth]{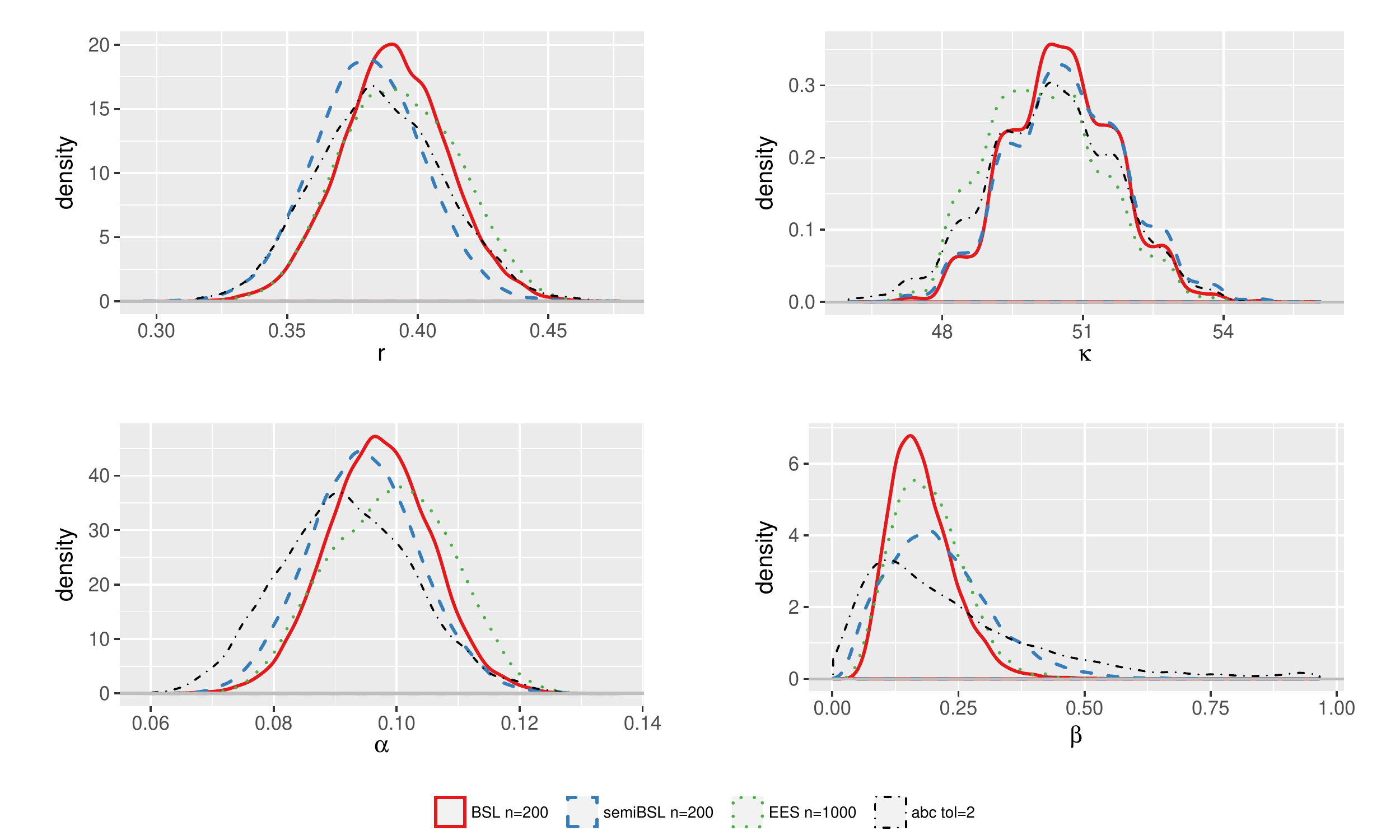}
    	\caption{Approximate marginal posterior distributions for the simple recruitment, boom and bust example. The tolerance used in MCMC ABC is $2$. The vertical lines indicate the true parameter values. The number of simulation used for each approach is given in the legend.}
    	\label{fig:post_bnb}
    \end{figure}

\begin{figure*}
		\centering
		\includegraphics[width=0.8\textwidth]{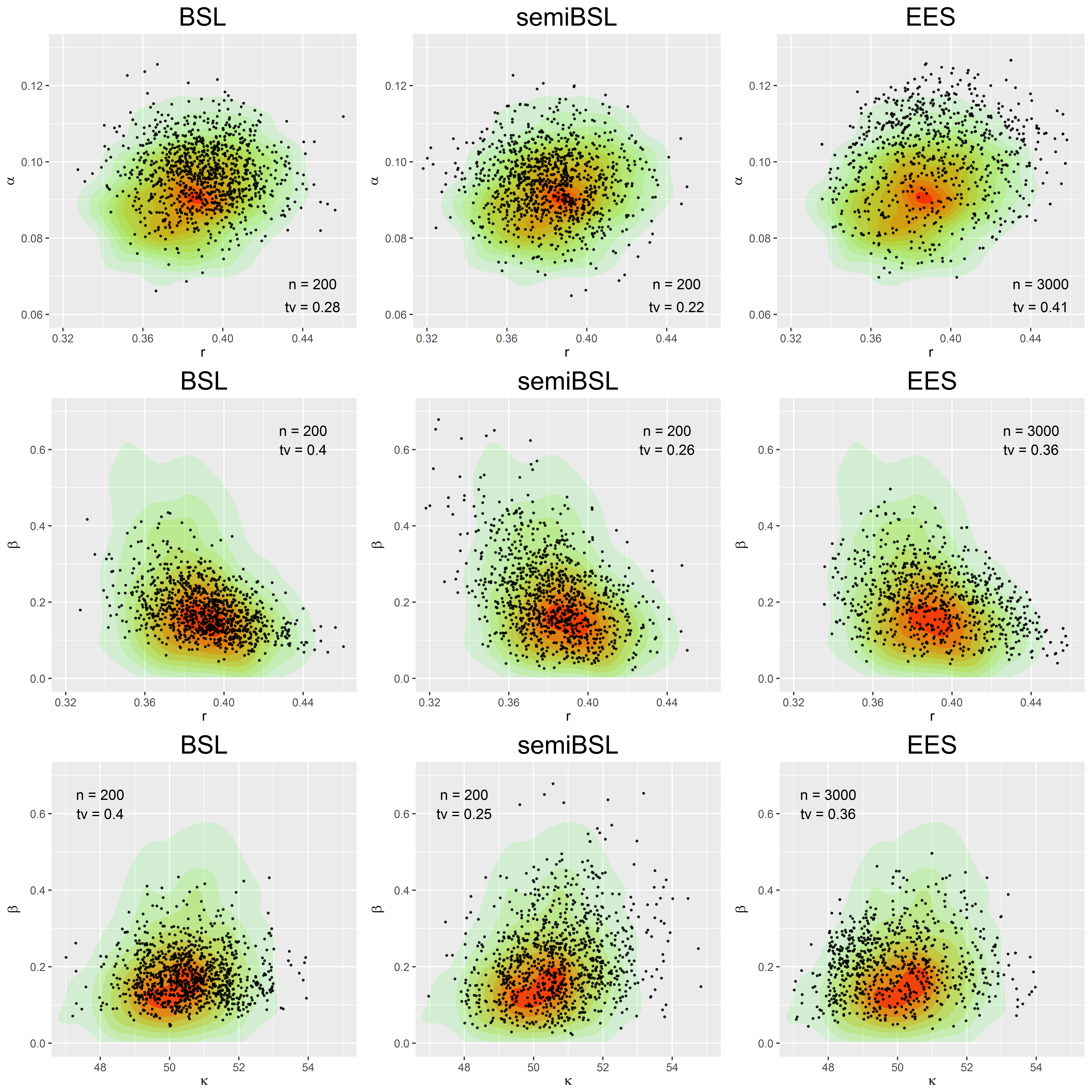}
		\caption{Bivariate scatter and contour plots of the posterior distributions of the simple recruitment, boom and bust example. The scatter plot is generated with thinned approximate sample obtained by BSL, semiBSL or EES methods. The contour plot is based on MCMC ABC with a tolerance of $2$. Only three pairs of the parameters are shown here, $r$ versus $\alpha$, $r$ versus $\beta$ and $\kappa$ versus $\beta$. The number of simulations per iteration $n$ and the total variation distance compared to the gold standard are shown in the corner of each plot.}
		\label{fig:scatter_bnb}
	\end{figure*}

	Distributions of the $12$-dimensional summary statistic (based on the true parameter value) are shown in Figure \ref{fig:summStat_bnb} in Appendix \ref{app:subsec:dist_summStat}. None of the chosen summary statistics are close to normal. Marginal posterior distributions by BSL, semiBSL, EES and ABC are shown in Figure \ref{fig:post_bnb}. The values of $n$ are given in the legend. With only four parameters and twelve summary statistics in this example, ABC can perform well and is treated as the gold standard.  In the MCMC ABC result, we manage to get $14.5$ thousand accepted samples out of 18 million iterations at tolerance $=2$. A Mahalanobis distance is used to compare summary statistics.  The covariance used in the Mahalanobis distance is taken to be the sample covariance of summary statistics independently generated from the model at $\vect{\theta} = (0.4,50,0.09,0.1)$. Figure \ref{fig:scatter_bnb} shows the bivariate scatterplot with overlaying contour plot (ABC result). It is evident that the semiBSL procedure is producing an approximation that is closer to ABC compared to BSL.  It does suggest that semiBSL is providing some robustness.  However, it is also evident that there is some difference between the ABC and semiBSL results. To gain some insights into the dependence structure between summaries, we consider bivariate scatterplots of the summaries, shown in Figure \ref{fig:summStat_corr_bnb} in Appendix \ref{app:subsec:scatter_summStat_bnb}. It is clear that there is a high degree of nonlinear dependence between many of the summaries, which cannot be captured by our Gaussian copula. Therefore this example is highly challenging for our semiBSL approach.
	
	\section{Discussion} \label{sec:conc}
	
	In this paper, we proposed a new method to relax the normality assumption in BSL and presented several examples of varying complexity to test the empirical performance of BSL and semiBSL. The new approach offered additional robustness in all of the considered examples.  Further, given the semi-parametric nature of the method, the computational gains of the fully parametric BSL are largely retained in terms of the number of model simulations required.  When model simulations per iteration required by BSL is non-negligible, then the additional cost incurred by semiBSL will be small.  Estimating marginal KDEs and the Gaussian rank correlation matrix is relatively straightforward.
	
	However, we did observe situations where standard BSL was remarkably robust to lack of normality, which is consistent with some previous literature including \citet{Price2018} and \citet{Everitt2017}.  Developing some theory around when we can expect standard BSL to work well would be useful.
	
	Previous BSL research \citep{Price2018} showed that the approximate posterior is very insensitive to the choice of $n$.  Surprisingly, we also found the semiBSL posterior to be relatively insensitive to $n$, albeit not as insensitive as BSL. We expect semiBSL to be less insensitive to $n$ since the choice of $n$ is more likely to impact kernel density estimates compared to the Gaussian synthetic likelihood. The sensitivity to $n$ results for semiBSL is presented in Appendix \ref{app:sensitivity2n}.
	
	The new approach was also compared with another robustified synthetic likelihood method, the EES \citep{Fasiolo2018}. Because of potential numerical issues with the standard empirical saddlepoint approximation, the EES has to resort to a tuning parameter called the decay, which shifts the estimation between a flexible saddlepoint one and a rigid normal distribution.  For roughly the same number of simulations, the posterior approximation by EES generally shows some improvements over BSL but less compared to semiBSL in the examples tested.
	
	Since we can obtain synthetic likelihood estimates of $-\infty$ with our approach for parameter values in the far tails of the posterior, we recommend that the practitioner firstly finds and initialises the MCMC at a parameter that produces a summary statistic distribution that has reasonable support for the observed statistic.  We note that standard BSL can also exhibit slow convergence when initialised in the tail of the posterior \citep{Price2018}.
	
	Note that we use an unrestricted correlation matrix $\vect{R}$ in equation \eqref{eq:pdf_semiBSL} throughout the main paper. However, it is possible to improve the computational efficiency with shrinkage estimation on $\vect{R}$. \citet{An2018} used the graphical lasso \citep{Friedman2008} in the standard BSL algorithm and proposed a novel penalty selection method so that the Markov chain has efficient mixing. We present the results using a straightforward shrinkage estimator proposed by \citet{Warton2008} in Appendix \ref{app:semiBSL_warton} and show a significant computational gain in the M/G/1 example.
	
	If the true underlying marginal distribution of a statistic is highly irregular, one limitation of our approach is that the number of simulations required for a KDE to capture this will be large. We point out that we only require estimates of the marginal distributions of the summary statistics at the observed statistic values, rather than estimating the entire marginal distribution. We note that future work could revolve around improving the performance of KDE. It could be beneficial to deliberately undersmooth in kernel estimation to reduce bias to get more accurate results in some cases. Another direction would be considering other adaptive kernel density estimation approaches, such as the balloon estimator and the sample point estimator \citep[e.g.][]{Terrell1992} , which may provide more stability and require less model simulations.
	
	There is still research to be done in improving the robustness of BSL.  Our semiBSL method relies on the Gaussian copula dependence structure. The results in the boom and bust example show a compromised performance of semiBSL when there exists strong nonlinear dependence structures between summary statistics. For future work, we plan to investigate other more flexible copula structures such as multivariate skew normal \citep{Sahu2008} and vine copulas \citep{Bedford2002}.  Another direction with great potential is to incorporate the semiBSL likelihood estimator into the variational Bayes synthetic likelihood approaches \citep[see][]{Ong2018b,Ong2018a} to speed up computation for a high dimensional statistic and/or parameter.

	Overall, we have demonstrated that our semiBSL approach can provide a significant amount of robustness relative to BSL with little or no additional computational cost in terms of the number of model simulations, while requiring no additional tuning parameters.
	
	\section*{Acknowledgements}
	CD was supported by an Australian Research Council's Discovery Early Career Researcher Award funding scheme (DE160100741).  ZA was supported by a scholarship under CDs Grant DE160100741 and a top-up scholarship from the Australian Research Council Centre of Excellence for Mathematical and Statistics Frontiers (ACEMS). DJN was supported by a Singapore Ministry of Education Academic Research Fund Tier 1 Grant (R-155-000-189-114). Computational resources and services used in this work were provided by the HPC and Research Support Group, Queensland University of Technology, Brisbane, Australia. The authors thank Alex Shestopaloff for sharing his code on exact MCMC for the M/G/1 model.
	
	\bibliographystyle{apalike} 
	\bibliography{refs}
	
	\newpage
	\begin{appendices}
		\section{Computational Efficiency Example} \label{app:computational_efficiency_example}
		
\begin{figure}[!htp]
			\centering
			\includegraphics[height=0.4\textheight,width=0.8\textwidth]{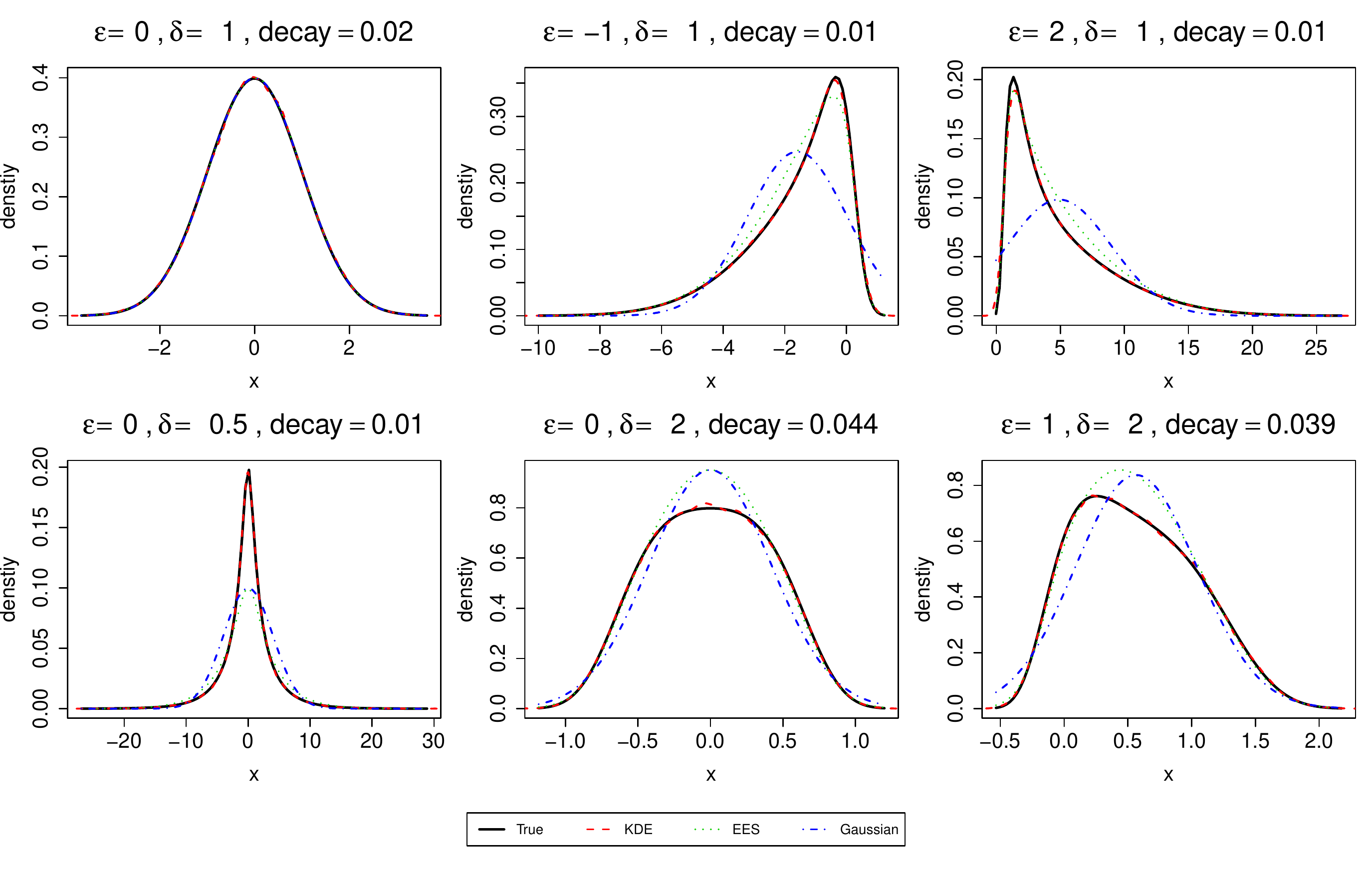}
			\caption{Univariate standard normal distribution transformed with the transformation in \eqref{eq:transformation} of the main paper applied for various combinations of $\epsilon$ and $\delta$.  The black solid line represents the true density. The rest indicate KDE, EES and Gaussian approximate distributions based on $1000$ simulations of the underlying distribution (see labels in the legend). The decay for EES is selected based on the $1000$ simulations (from left to right, top to bottom, the decay values are 0.02, 0.01, 0.01, 0.01, 0.044 and 0.039).}
			\label{fig:summStat_ma2}
			%		\caption{Distribution of the summary statistics of the MA(2) example. The solid black lines represent the distributions of $y_1$ using different pairs of transformation parameter $\epsilon$ and $\delta$. The overlaid red dashed lines are the Gaussian densities computed with sample mean and covariance.}
		\end{figure}

		Here we provide a simple example to gain some insight into the computational efficiency of the semi-parametric estimator compared to the standard synthetic likelihood.  We assume that observed data have been drawn from a Gaussian, $\vect{\tilde{y}} \sim \mathcal{N}(\vect{\mu},\vect{\Sigma})$ where $\vect{\tilde{y}} \in \mathbb{R}^d$ and $d$ is the dimension of $\vect{\tilde{y}}$.  To disturb the normality assumption we consider the same transformation used in section \ref{subsec:ma2} in the main paper. We take the transformed data as our summary statistic.  Figure \ref{fig:summStat_ma2} shows the marginal distribution of any individual data point with six pairs of $\epsilon$ and $\delta$. Note that $\epsilon=0$ and $ \delta=1$ corresponds to no transformation. Based on Figure \ref{fig:summStat_ma2}, the EES seems to provide a reasonable approximation in the presence of significant skewness, but performs less well when there is significant kurtosis.  Note that these results are based on $n=1000$ simulations from the underlying density, producing relatively small decay values.  Despite this, the EES approximation does not perform as well as KDE for this simulation size.
		
		%To disturb the normality assumption we consider the following transformation of each data component
		%\begin{align}
		%y_i &= \sinh \Big(\dfrac{1}{\delta} \sinh^{-1}(\tilde{y}_i + \epsilon) \Big), \mbox{  for } i=1,\ldots,d,
		%\label{eq:transformation}
		%\end{align}
		%where $\epsilon$ and $\delta$ control the level of skewness and kurtosis, respectively. 

		Here we compute 500 independent estimates of the semi-parametric and standard log synthetic likelihoods with observed datasets generated using $\vect{\mu} = \vect{0}$ and where the $(i,j)$th element of $\vect{\Sigma}$ is given by $\Sigma_{i,j} = 0.5^{|i-j|}$.  We consider all combinations of $d=20, 50, 100$ and $\delta = 0.5, 0.7, 1, 2, 3$, with $\epsilon = 0$ in all cases.  Values of $\delta=2, 3$ and $\delta=0.5, 0.7$ indicate tails lighter and heavier than the normal distribution, respectively.  The value $\delta=0.5$ produces a very heavy tailed distribution.  
		
		It is easy to show that the true log-likelihood for a given $\vect{y}$ is given by
		\begin{align*}
		\log p(\vect{y}|\vect{\mu}, \vect{\Sigma}) &= \log \mathcal{N}(\vect{\tilde{y}};\vect{\mu},\vect{\Sigma}) + \sum_{i=1}^d \left\{ \log\left(\cosh(\delta \sinh^{-1}(y_i)-\epsilon)\right) + \log\delta - \frac{1}{2}\log(1+y_i^2)\right\},
		\end{align*}
		where the second term is the log of the Jacobian of the transformation.  
		
		%	For a given $\vect{y}$, the likelihood is always evaluated/approximated at the given values of $\vect{\mu}$ and $\vect{\Sigma}$ above.
		
\begin{figure}[!htp]
			\centering
			\includegraphics[width=0.95\textwidth]{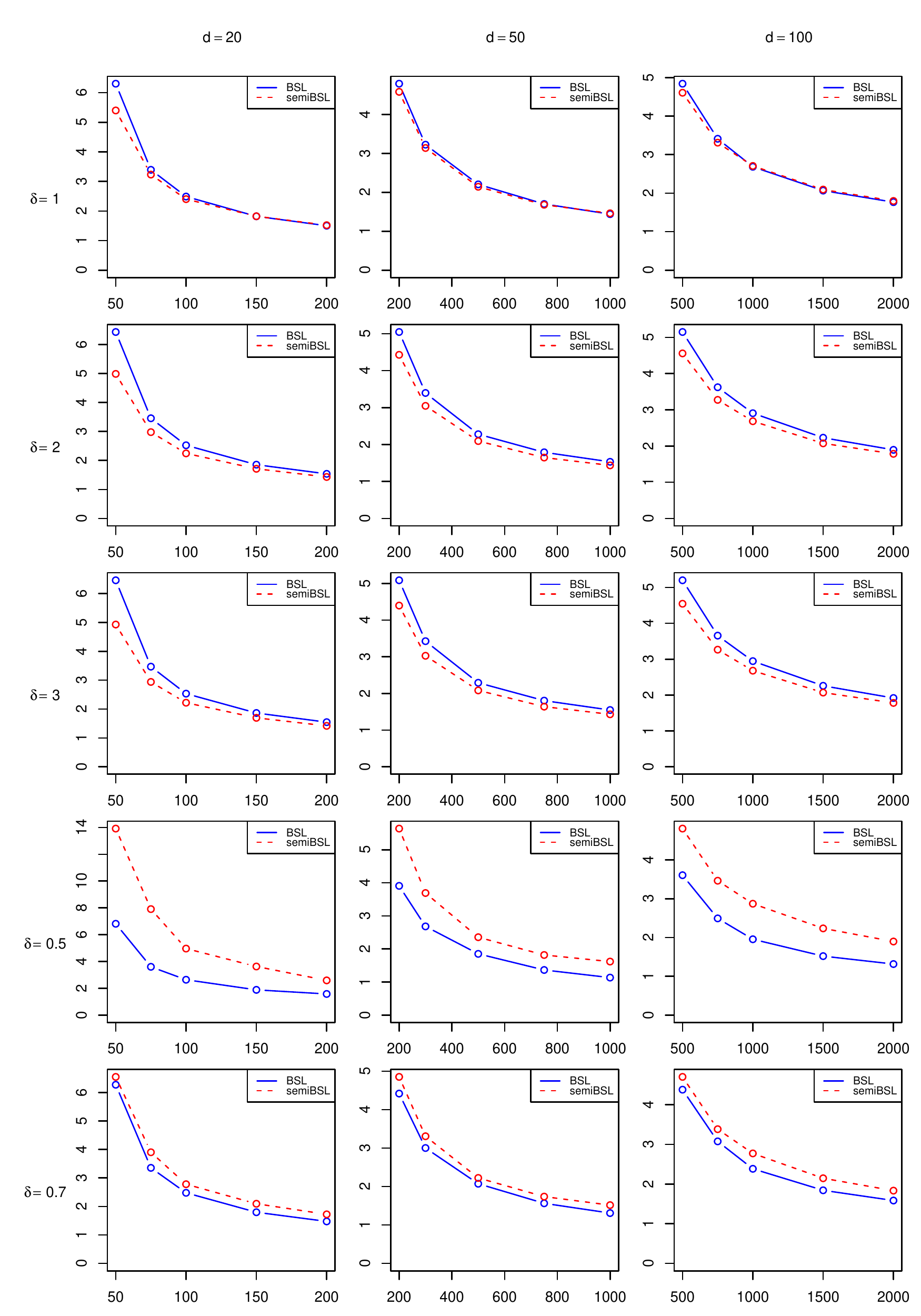}
			\caption{Estimated standard deviations of the semi-parametric (red dash) and standard (blue solid) log synthetic likelihoods for the toy example.  The x-axis denotes the number of simulations $n$ used to estimate the log-likelihood and the y-axis denotes the estimated standard deviation of the estimated log-likelihood. Columns from left to right are for $d=20, 50$ and $100$, respectively.}
			\label{fig:toy_std}
		\end{figure}

\begin{figure}[!htp]
			\centering
			\includegraphics[width=0.95\textwidth]{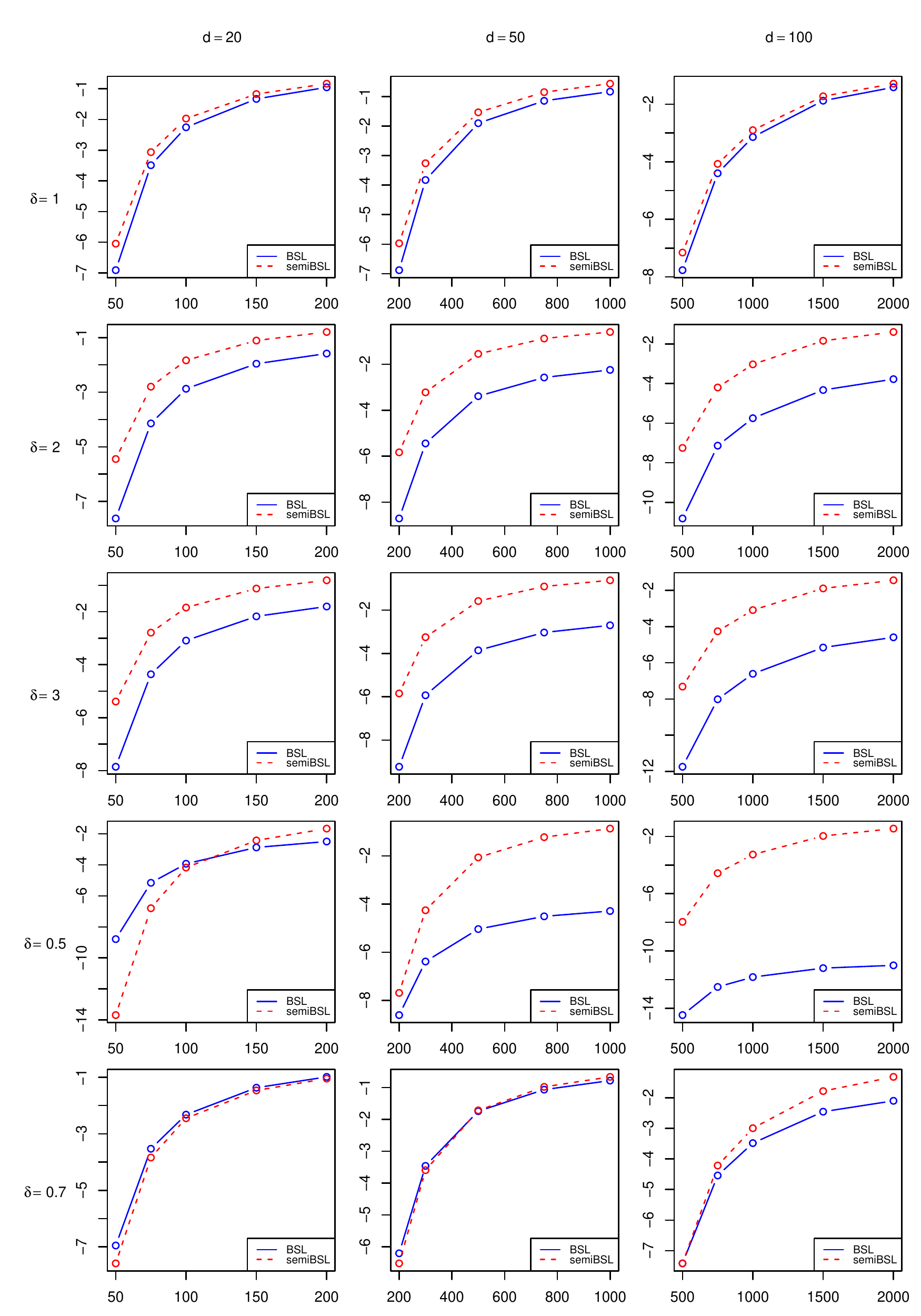}
			\caption{Estimated bias of the semi-parametric (red dash) and standard (blue solid) log synthetic likelihoods for the toy example.  The x-axis denotes the number of simulations $n$ used to estimate the log-likelihood and the y-axis denotes the estimated bias of the estimated log-likelihood. Columns from left to right are for $d=20, 50$ and $100$, respectively.}
			\label{fig:toy_bias}
		\end{figure}

		The estimated standard deviation and bias of the estimated synthetic likelihoods (both semi-parametric and standard approach) for different combinations of $\delta$ and $n$ (the number of simulations used in the synthetic likelihood estimator) are shown in Figure \ref{fig:toy_std} and Figure \ref{fig:toy_bias}, respectively.  From row 1 of these figures, it is surprising that the semi-parametric estimator performs almost as well as the synthetic likelihood despite the normality assumption being perfect ($\delta=1$).  We are effectively losing nothing using nonparametric kernel estimates of the marginals instead of the true normal marginals.  It is likely that the variance of the standard synthetic likelihood estimator is dominated by having to estimate a high-dimensional covariance matrix. 
		
		The next two rows of Figure \ref{fig:toy_std} and Figure \ref{fig:toy_bias} are for when the true marginals have lighter tails than the normal distribution.  It is interesting that the semi-parametric approach gives a smaller variance and bias compared to the standard synthetic likelihood.  The lack of normality is contributing to the relative poor performance of the standard synthetic likelihood estimator.
		
		The final two rows of Figure \ref{fig:toy_std} and Figure \ref{fig:toy_bias} are for when the true marginals have heavier tails than the normal distribution.  Despite the lack of normality, the standard estimator appears to have a smaller variance compared to the semi-parametric approach.  Upon further investigation, we find that the high variance is due to the fact that when data is generated with heavy tails, an observed component of the data can fall out in the tail of the distribution.  It can fall so far out that all simulated values are either less than or greater than the observed component.  Given that kernels used in kernel density estimation do not have heavy tails, the marginal density based on the KDE can be severely underestimated, and numerically equal to $-\infty$, as occurred for $d=20$ and $n=75$ in the bottom left panel of Figure \ref{fig:toy_std}.  This issue is generally alleviated with larger $n$, but can still inflate the variance.  It is important to note that the odd occurrence of massively underestimated log-likelihoods will have limited impact on the computational efficiency of BSL.  These will simply be rejected by the MCMC algorithm; it is the overestimated log-likelihoods that substantially decrease efficiency as it can lead to sticky behaviour in the MCMC chain.

		This simple example demonstrates that the semi-parametric estimator can still maintain the computational benefits that BSL has over ABC.  The semiBSL approach may even be more computationally efficient than BSL in situations of non-normality.
		%However, we are yet to explore the impact of non-normality on the quality of posterior approximations.  We aim to do that in the next section.

		\newpage
		\section{Extra Results} \label{app:extra_figures}
		
		In this appendix, for each example, we show the distribution of the summary statistics and also the estimated univariate posterior distributions obtained by the different approximations and the gold standard discussed in the main paper.
		
		\subsection{Distributions of the Summary Statistics} \label{app:subsec:dist_summStat}
		
		We present some of the marginal distributions of summary statistics in each example to give an indication of the non-normalities that can be encountered. For these illustrations, the marginal distributions are approximated with KDE based on a sufficiently large number of simulations at either the true value or a point estimate of $\vect{\theta}$ with reasonable posterior support. 
		
\begin{figure}[!htp]
			\centering
			\includegraphics[width=0.8\textwidth]{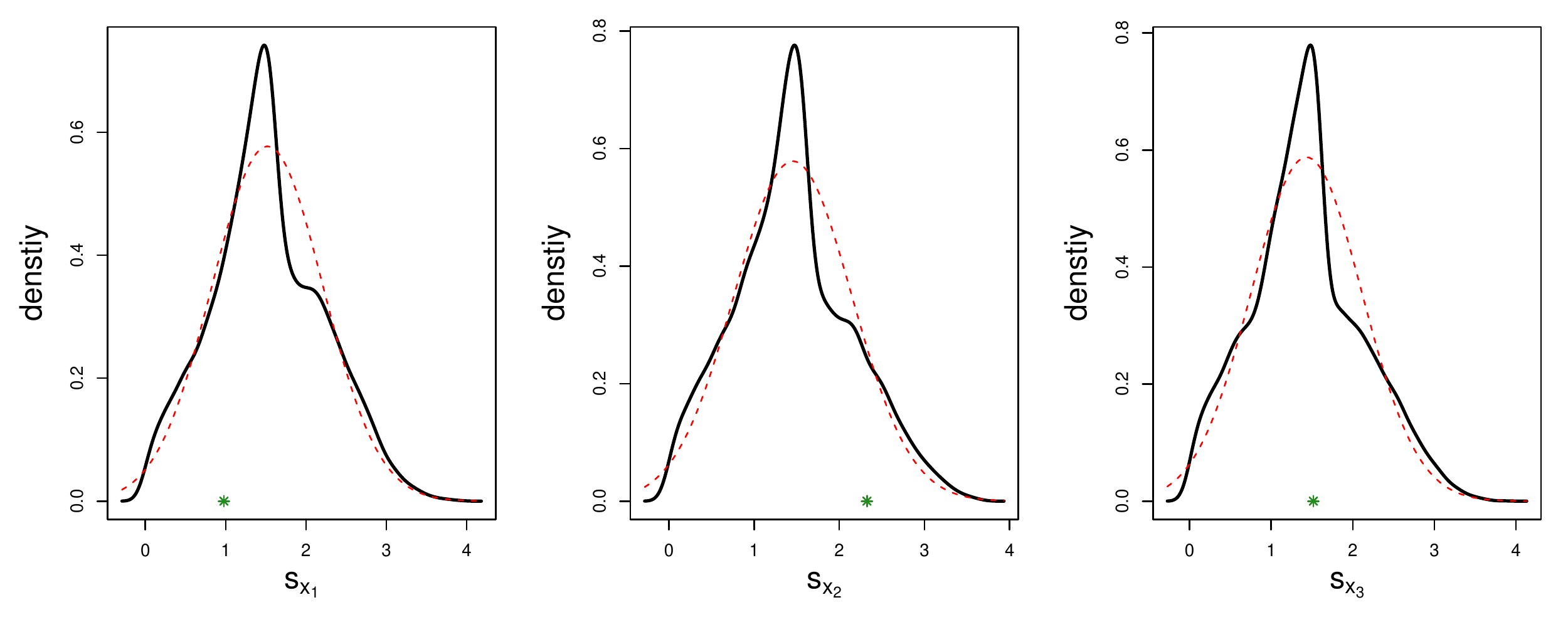}
			\caption{Distribution of the summary statistics of the M/G/1 example. The solid black line indicates the KDE with a large number of simulations at the true of $\vect{\theta}$. The red dashed line is given by a Gaussian approximation using the sample mean and variance. Location of the observed summary statistics is marked on x-axis in dark green.}
			\label{fig:summStat_mg1}
		\end{figure}

\begin{figure}[!htp]
			\centering
			\includegraphics[width=0.8\textwidth]{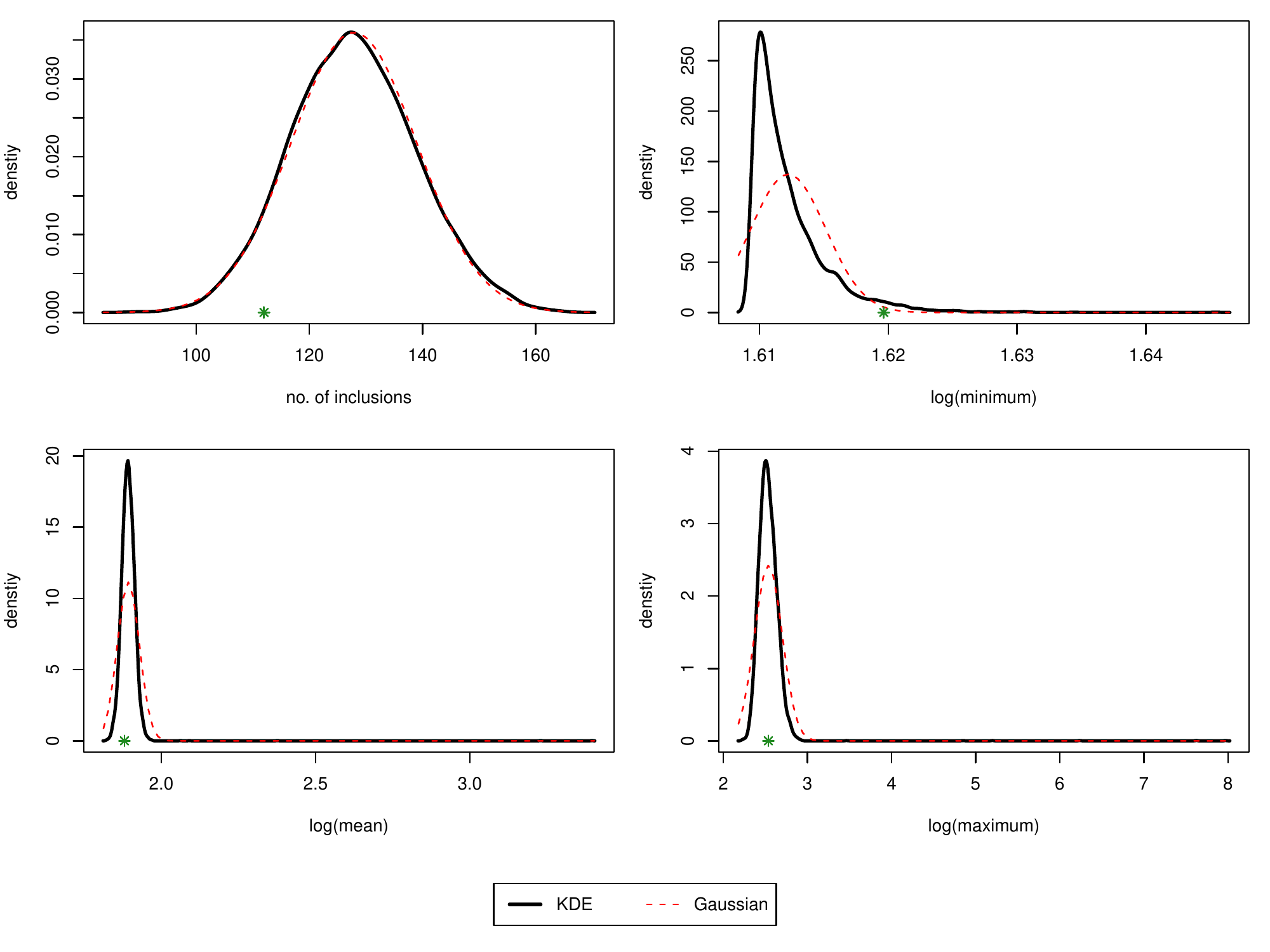}
			\caption{Distribution of the summary statistics of the stereological extreme example. The solid black line indicates the KDE with a large number of simulations at a point estimate of $\vect{\theta}$. The red dashed line is given by a Gaussian approximation using the sample mean and variance. Location of the observed summary statistics is marked on x-axis in dark green.}
			\label{fig:summStat_stereo}
		\end{figure}

\begin{figure}[!htp]
			\centering
			\includegraphics[width=0.8\textwidth]{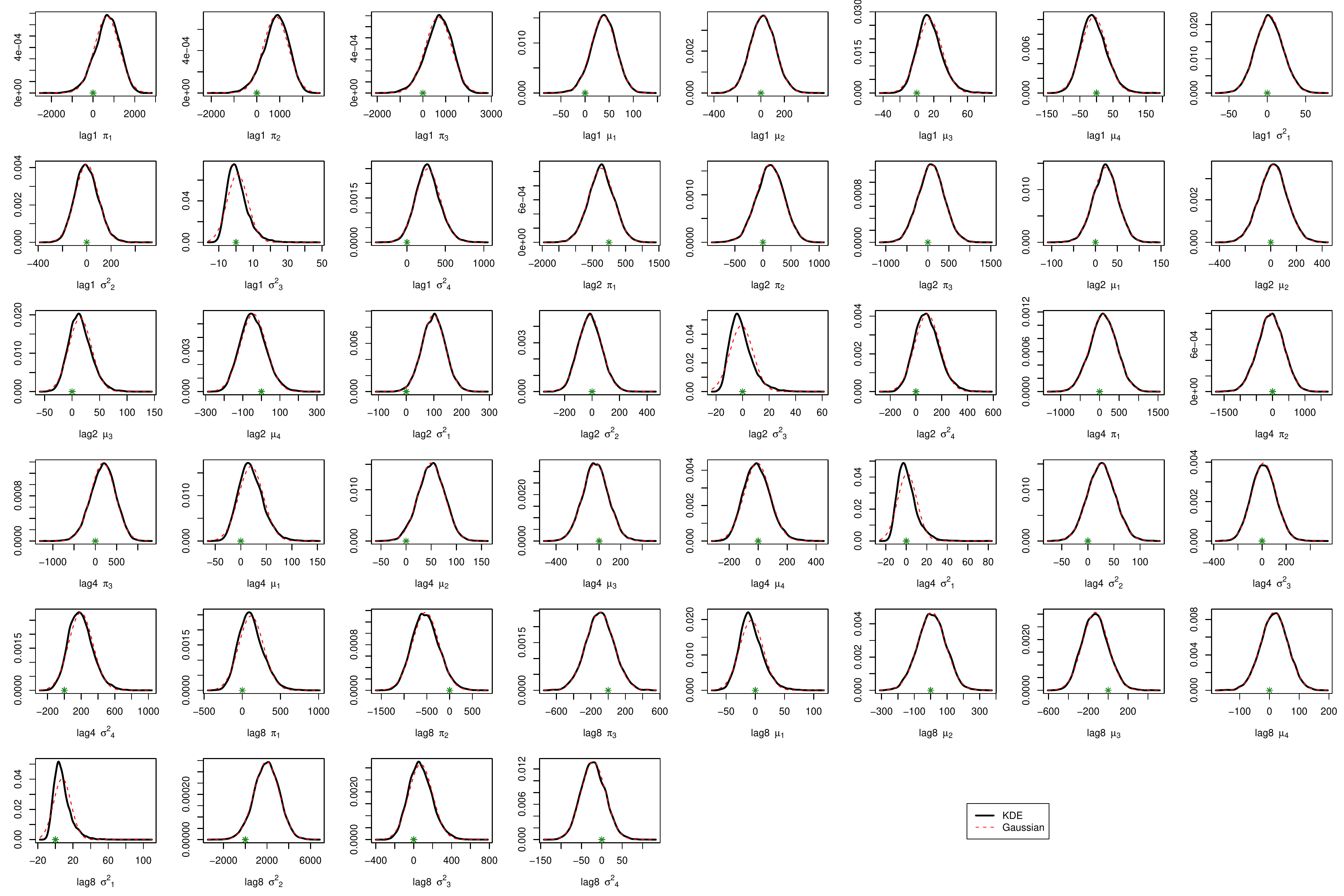}
			\caption{Distribution of the score summary statistics of the Fowler's toads example. The solid black line indicates the KDE from a large number of simulations at the true $\vect{\theta}$. The red dashed line is given by a Gaussian approximation using the sample mean and variance. Location of the observed summary statistics is marked on x-axis in dark green.}
			\label{fig:summStat_scores_toads}
		\end{figure}

\begin{figure}[!htp]
			\centering
			\includegraphics[width=0.8\textwidth]{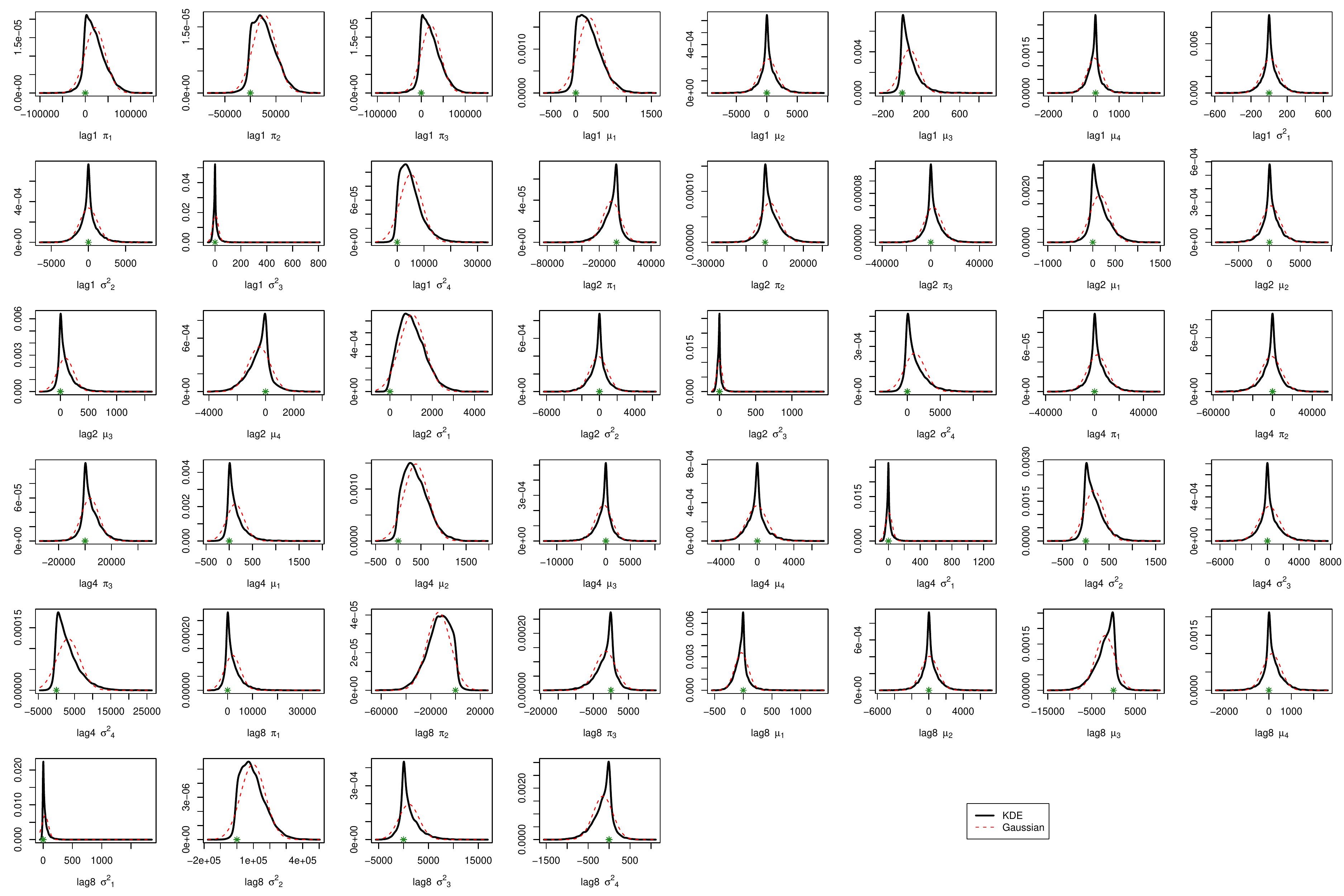}
			\caption{Distribution of the transformed summary statistics (power=$1.5$) of the Fowler's toads example. The solid black line indicates the KDE from a large number of simulations at the true $\vect{\theta}$. The red dashed line is given by a Gaussian approximation using the sample mean and variance. Location of the observed summary statistics is marked on x-axis in dark green.}
			\label{fig:summStat_power1.5_toads}
		\end{figure}

\begin{figure}[!htp]
			\centering
			\includegraphics[width=0.8\textwidth]{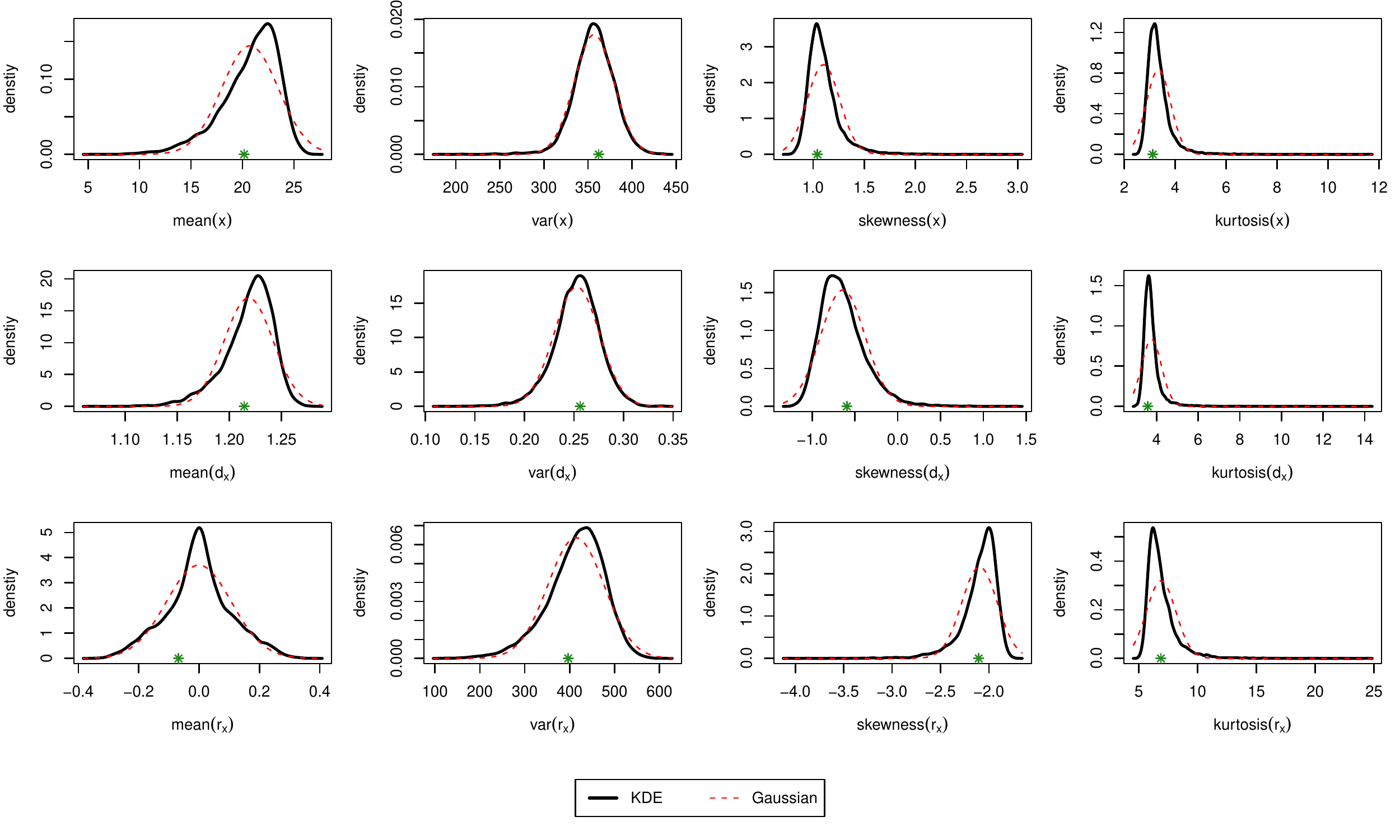}
			\caption{Distribution of the summary statistics of the simple recruitment, boom and bust example. The solid black line indicates the KDE with a large number of simulations at a point estimate of $\vect{\theta}$. The red dashed line is given by a Gaussian approximation using the sample mean and variance. Location of the observed summary statistics is marked on x-axis in dark green.}
			\label{fig:summStat_bnb}
		\end{figure}

		For the MA(2) example, the distributions of summary statistics is similar to Figure \ref{fig:summStat_ma2} in Appendix \ref{app:computational_efficiency_example}. Figure \ref{fig:summStat_mg1} shows the distributions of the first three summary statistics of the M/G/1 example. Figure \ref{fig:summStat_stereo} shows the distributions of the summary statistics of the stereological extreme example. Two figures are provided for the Fowler's toads example, Figure \ref{fig:summStat_scores_toads} (using the original score summary statistics) and Figure \ref{fig:summStat_power1.5_toads} (using $p=1.5$ in the transformation of summary statistics). It is clear that the transformation can disturb the normality significantly in this example. Distributions of the summary statistics for the boom and bust example are presented in Figure \ref{fig:summStat_bnb}.
		
		%Figure \ref{fig:summStat_mg1} shows the distributions of the first three summary statistics based on the true parameter value, which are similar to the rest as the queue has a steady state and little transient behaviour. The distribution has an interesting and uncommon non-symmetric curve. As can be seen in the figure, the synthetic likelihood does not approximate well near the sharp mode.
		
		\newpage
		\subsection{Distributions of the Approximate Marginal Posteriors} \label{app:subsec:dist_maginal_post}
		
\begin{figure}[!htp]
			\centering
			\includegraphics[width=0.9\textwidth]{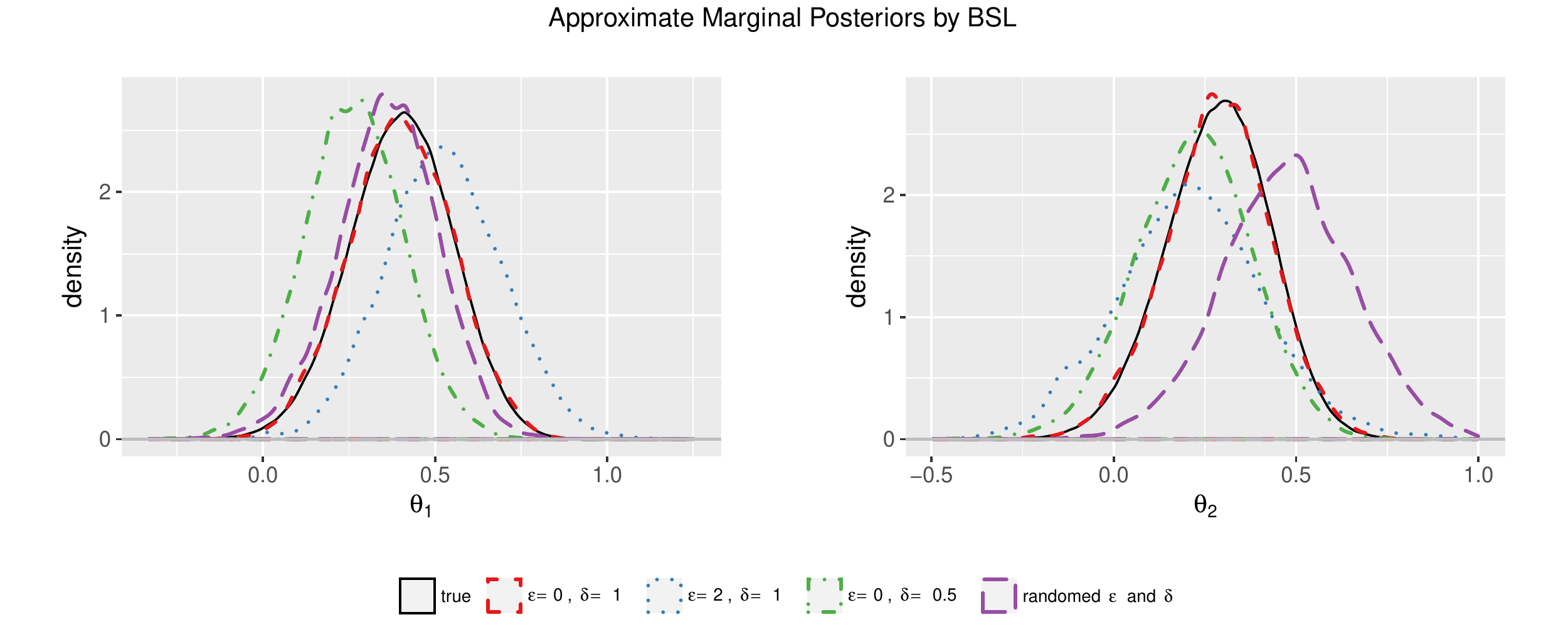}\\
			\vspace{0.5cm}
			\includegraphics[width=0.9\textwidth]{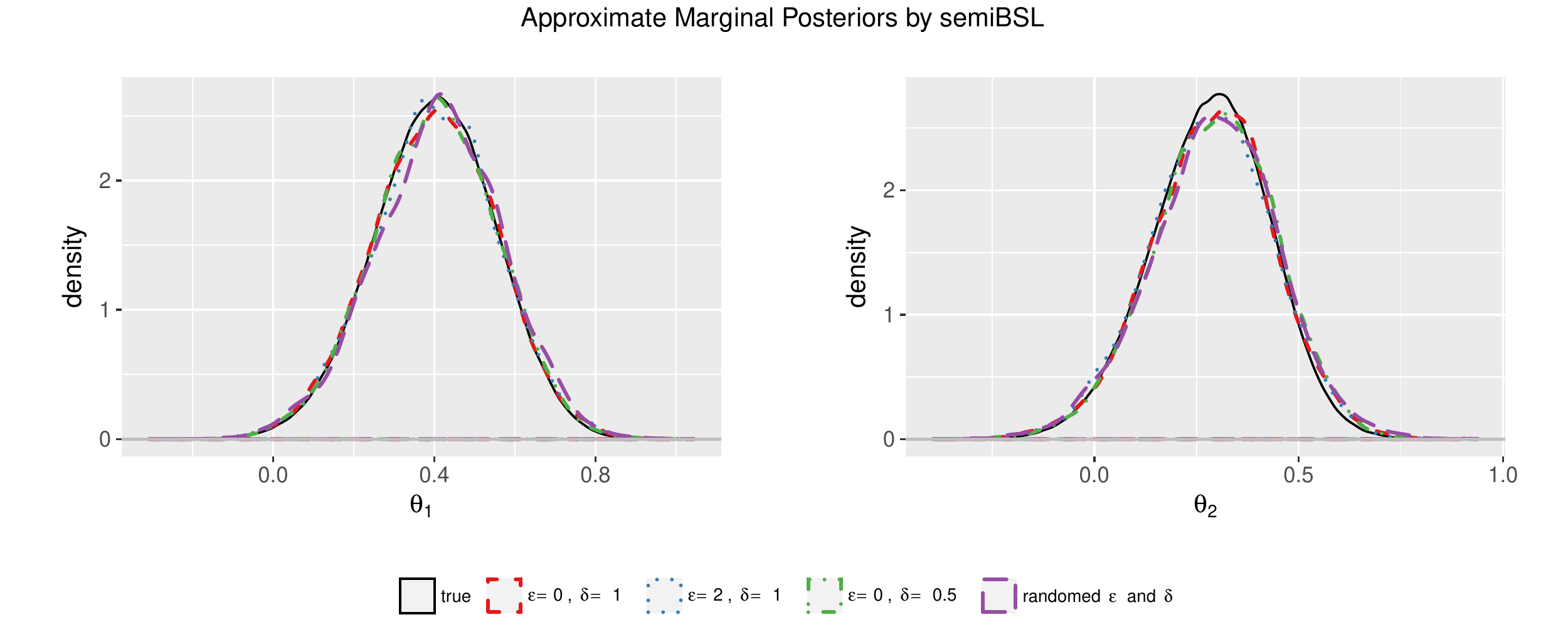}\\
			\vspace{0.5cm}
			\includegraphics[width=0.9\textwidth]{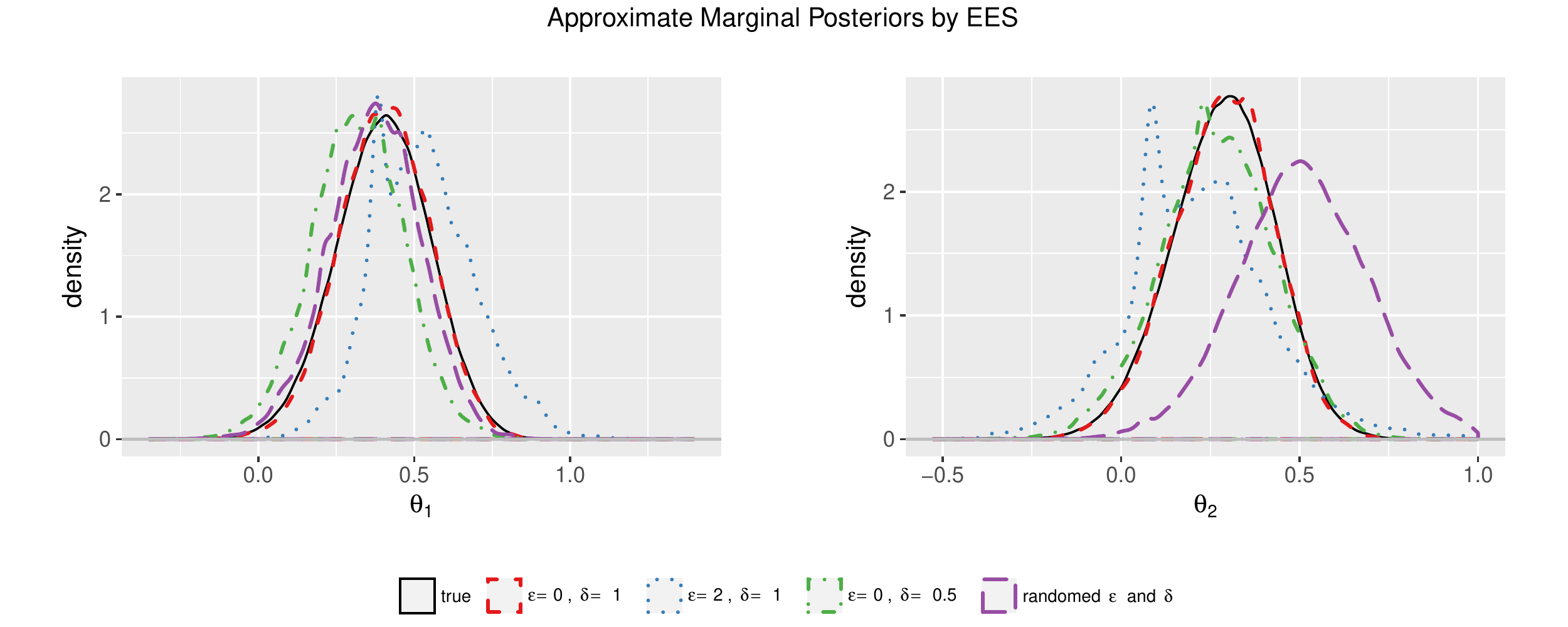}
			\caption{Estimated marginal approximate posteriors (BSL in the first row, semiBSL in the second row and EES in the third row) for the MA(2) example using different pairs of transformation parameters for the summary statistics.}
			\label{fig:post_ma2}
		\end{figure}

\begin{figure}[!htp]
			\centering
			\includegraphics[width=1\textwidth]{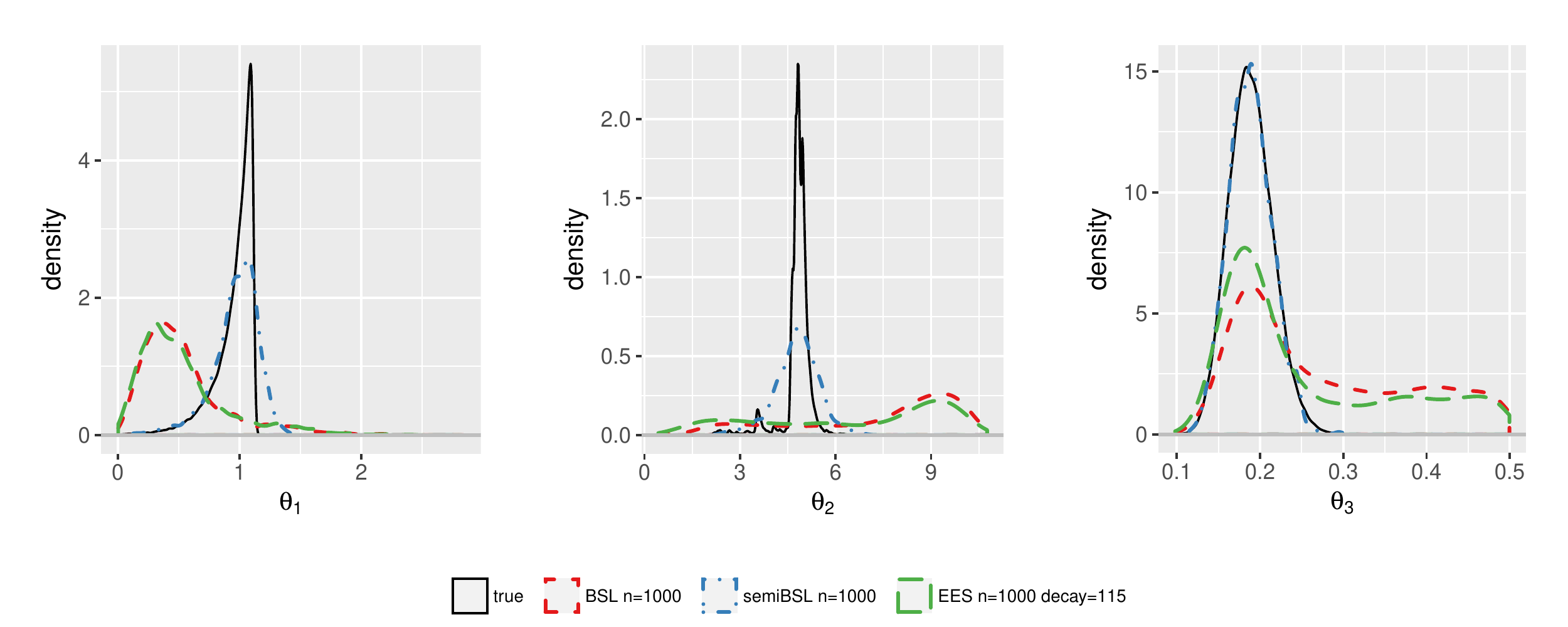}
			\caption{Approximate marginal posterior distributions of the M/G/1 example. The solid black line shows the ``true'' posterior distribution. The number of simulation used for each approach is given in the legend.}
			\label{fig:post_mg1}
		\end{figure}

\begin{figure}[!htp]
			\centering
			\includegraphics[width=1\textwidth]{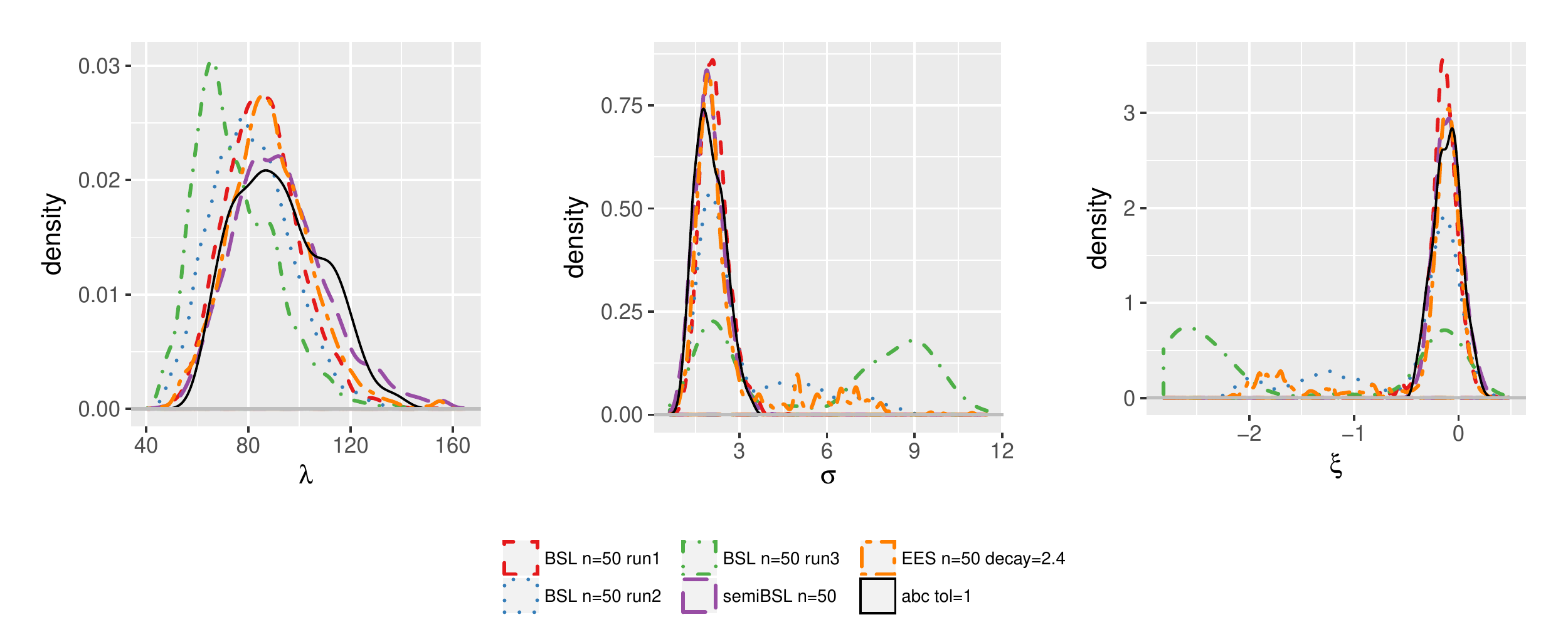}
			\caption{Approximate marginal posterior distributions of the stereological extreme example. The number of simulation used for each approach is given in the legend.}
			\label{fig:post_stereo}
		\end{figure}

\begin{figure}[!htp]
			\centering
			\includegraphics[width=0.8\textwidth]{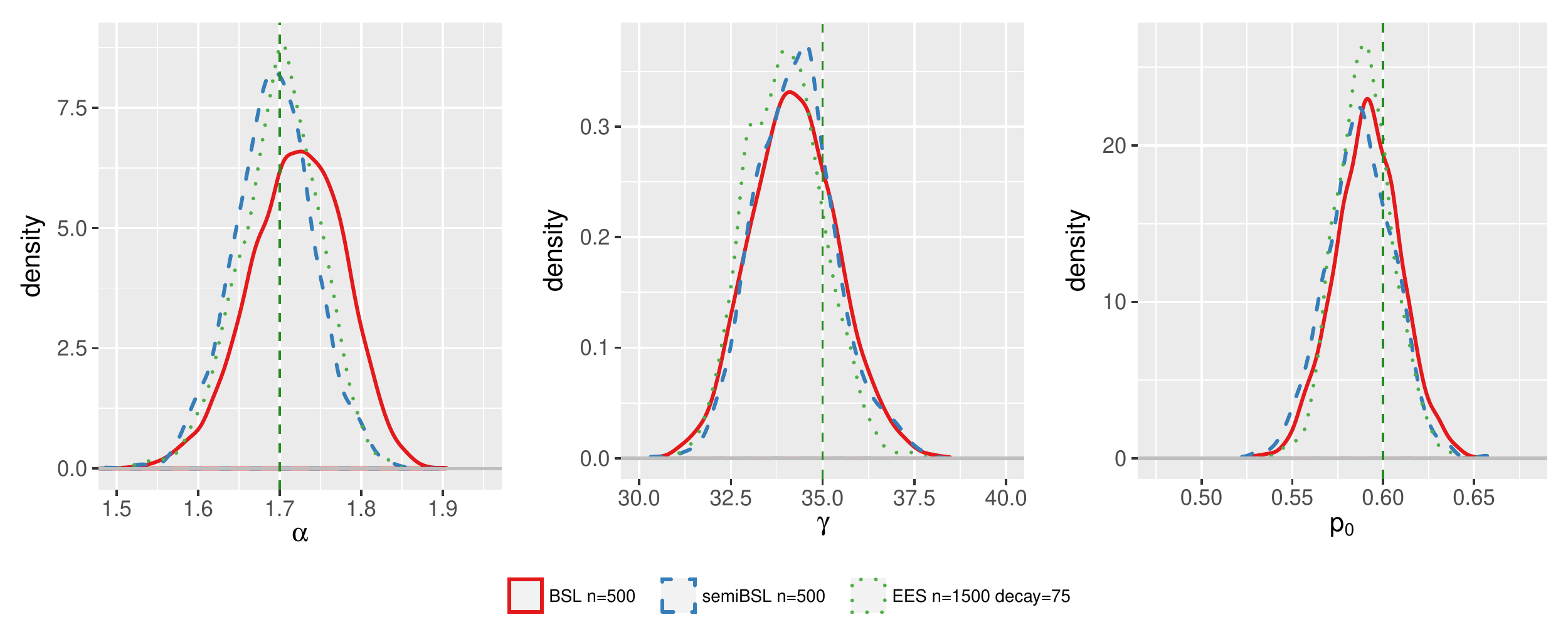}
			\includegraphics[width=0.8\textwidth]{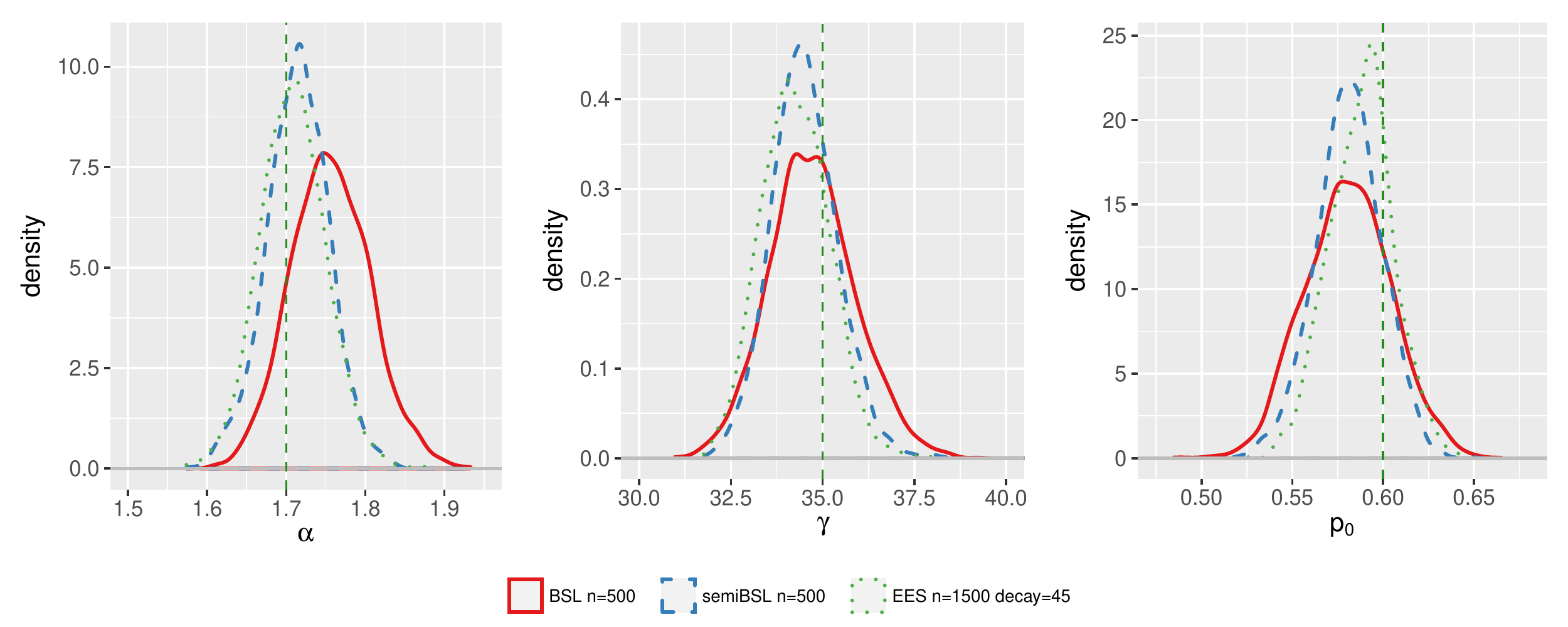}
			\includegraphics[width=0.8\textwidth]{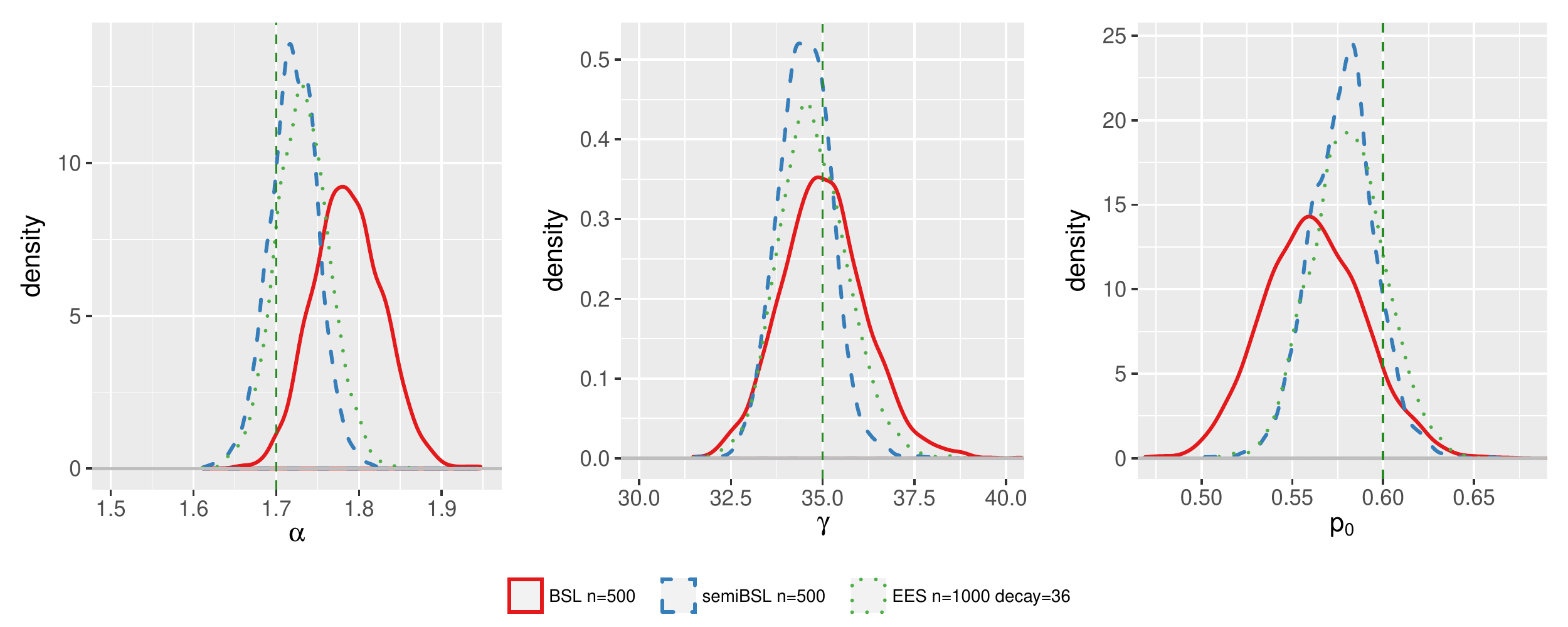}
			\caption{Approximate marginal posterior distributions by BSL, semiBSL and EES of the Fowler's toads example using scores (first row), transformed scores with $p=1.5$ (second row) and transformed scores with $p=2$ (third row). The vertical line indicate the true parameter value.}
			\label{fig:post_toads}
		\end{figure}

		Figure \ref{fig:post_ma2}, \ref{fig:post_mg1} and \ref{fig:post_stereo} show the approximate marginal posterior distributions for the MA(2), M/G/1 and stereological extreme example, respectively. Figure \ref{fig:post_toads} compares the approximate marginal posteriors obtained by BSL, semiBSL and EES. The results are consistent with the interpretations provided in the main paper.
		
		%Figure \ref{fig:post_stereo} shows the marginal posterior distributions with three BSL runs, one semiBSL run and one EES run. Note that the three BSL posteriors were generated with identical input arguments, however the results differ substantially. On the contrary, semiBSL results are consistent with multiple runs (results not shown). It is worth pointing out that MCMC BSL is having trouble exploring the full parameter space due to the rarely occurring outliers, but the current result is enough to show the difference in posterior approximation.
		
		\newpage
		\subsection{Scatterplot of the Bivariate Summary Statistics of the Boom and Bust Example} \label{app:subsec:scatter_summStat_bnb}
		
\begin{figure}[!htp]
			\centering
			\includegraphics[width=0.8\textwidth]{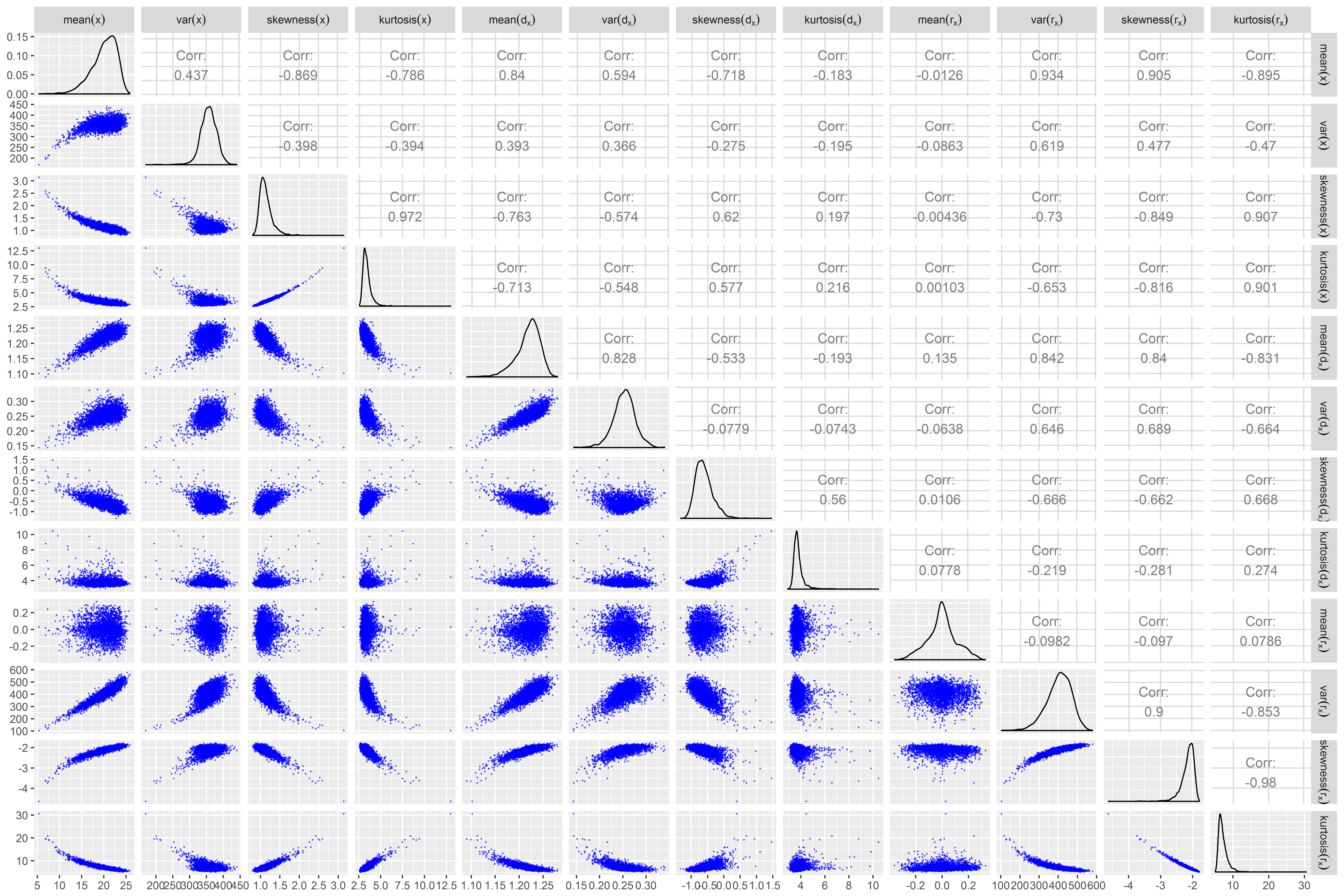}
			\caption{Bivariate scatterplots of the summary statistics of the simple recruitment, boom and bust example.}
			\label{fig:summStat_corr_bnb}
		\end{figure}

		As mentioned in the main paper, the boom and bust example is particularly challenging for the semiBSL approach due to the strong nonlinear dependence structure between the summary statistics. Figure \ref{fig:summStat_corr_bnb} shows the bivariate scatterplots of the summary statistics simulated with $\vect{\theta} = (0.4,50,0.09,0.1)$.
		
		\newpage
		\section{Sensitivity to $n$} \label{app:sensitivity2n}
		
		For brevity of the main paper, we only present results for the value of $n$ (number of model simulations per iteration) that maximise the computational efficiency.  Here we explore whether the results for semiBSL are sensitive to $n$. \citet{Price2018} show that the standard BSL method is remarkablely insensitive to $n$. The property also carries over to BSLasso, a relevant BSL approach with graphical lasso shrinkage estimation on the inverse covariance matrix.
		
\begin{figure}[!htp]
			\centering
			\includegraphics[width=0.95\textwidth]{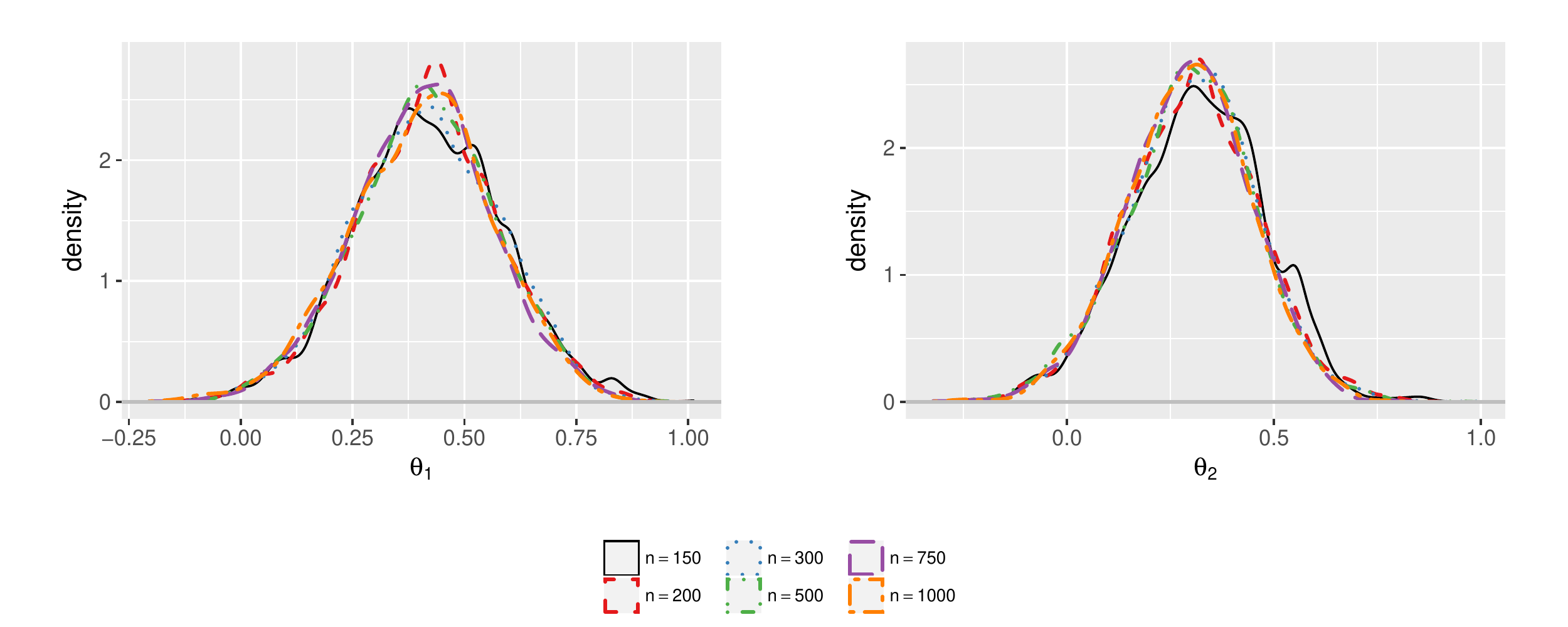}
			\caption{Approximate marginal posterior distributions for the MA(2) example with various number of simulations per iteration. Transformation parameters for the summary statistics is randomed.}
			\label{fig:sens_ma2}
		\end{figure}

\begin{figure}[!htp]
			\centering
			\includegraphics[width=0.95\textwidth]{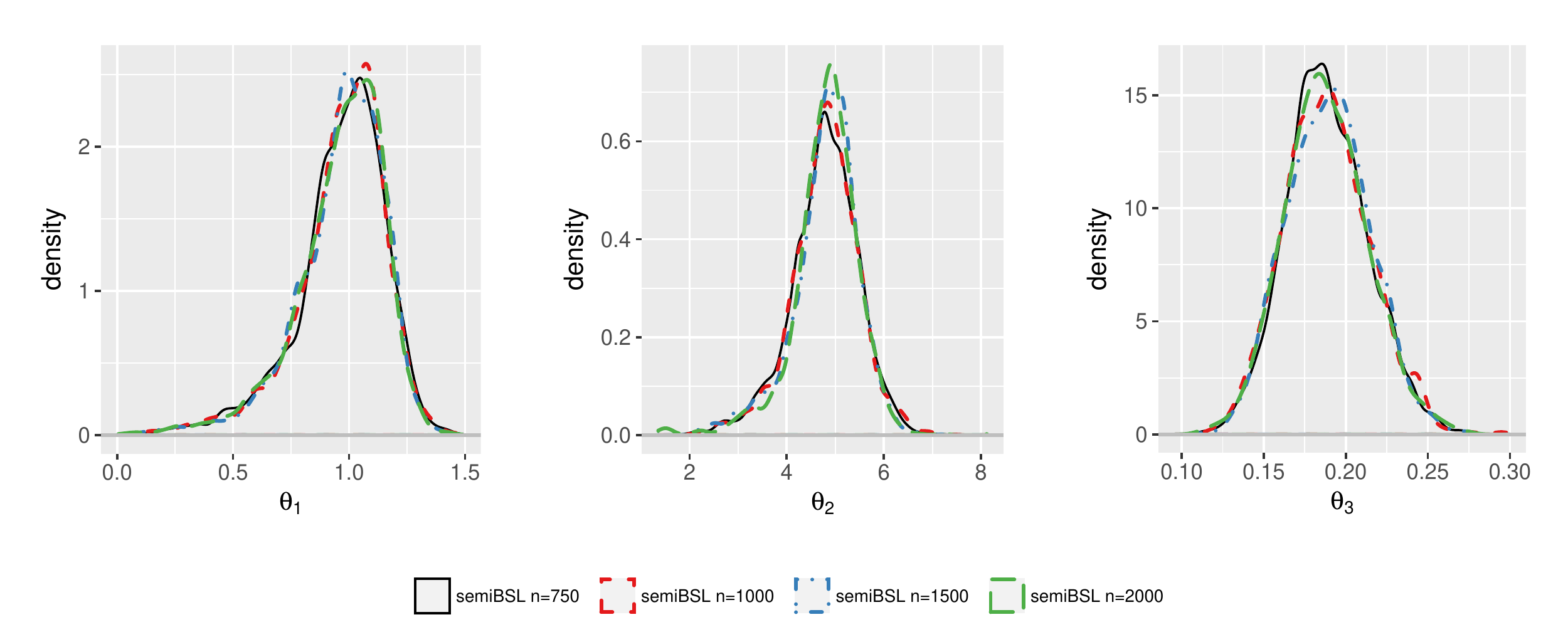}
			\caption{Approximate marginal posterior distributions for the M/G/1 example with various number of simulations per iteration.}
			\label{fig:sens_mg1}
		\end{figure}

\begin{figure}[!htp]
			\centering
			\includegraphics[width=0.95\textwidth]{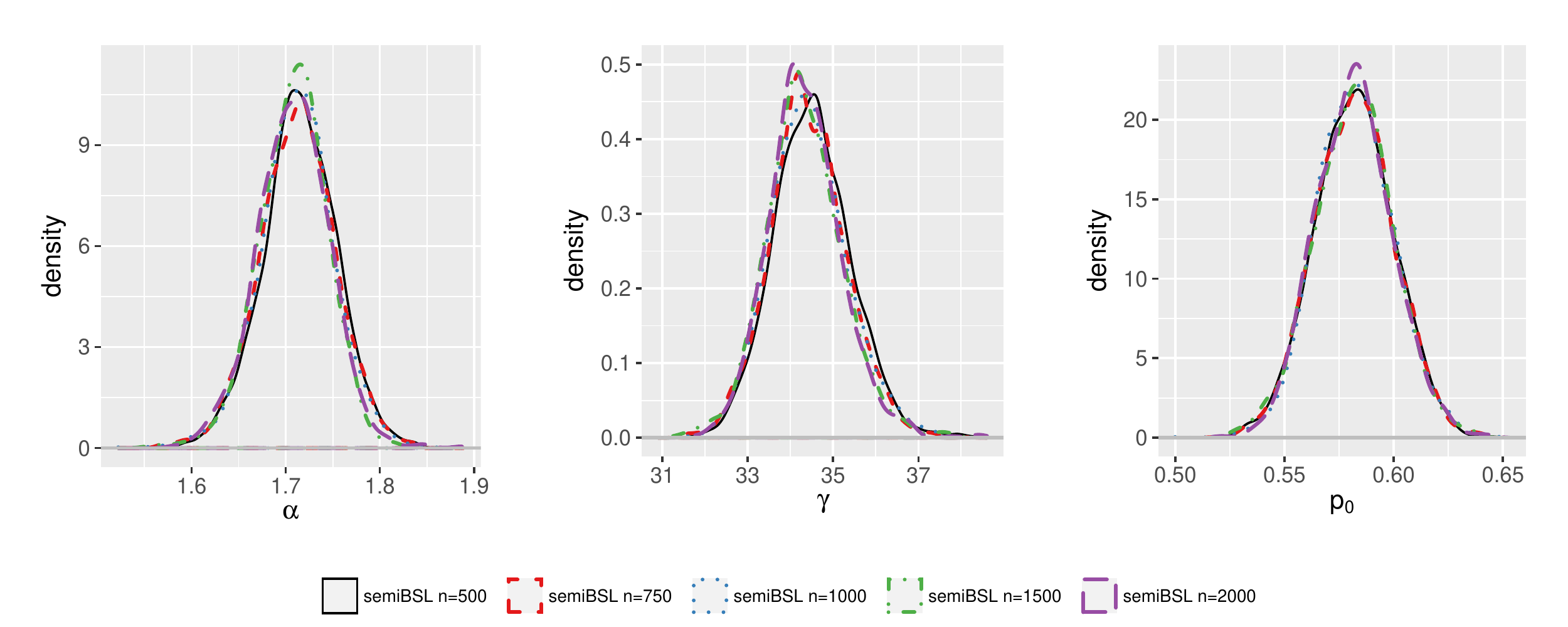}
			\caption{Approximate marginal posterior distributions for the toads example with various number of simulations per iteration. The Transformation power is $1.5$.}
			\label{fig:sens_toads}
		\end{figure}

\begin{figure}[!htp]
			\centering
			\includegraphics[width=0.95\textwidth]{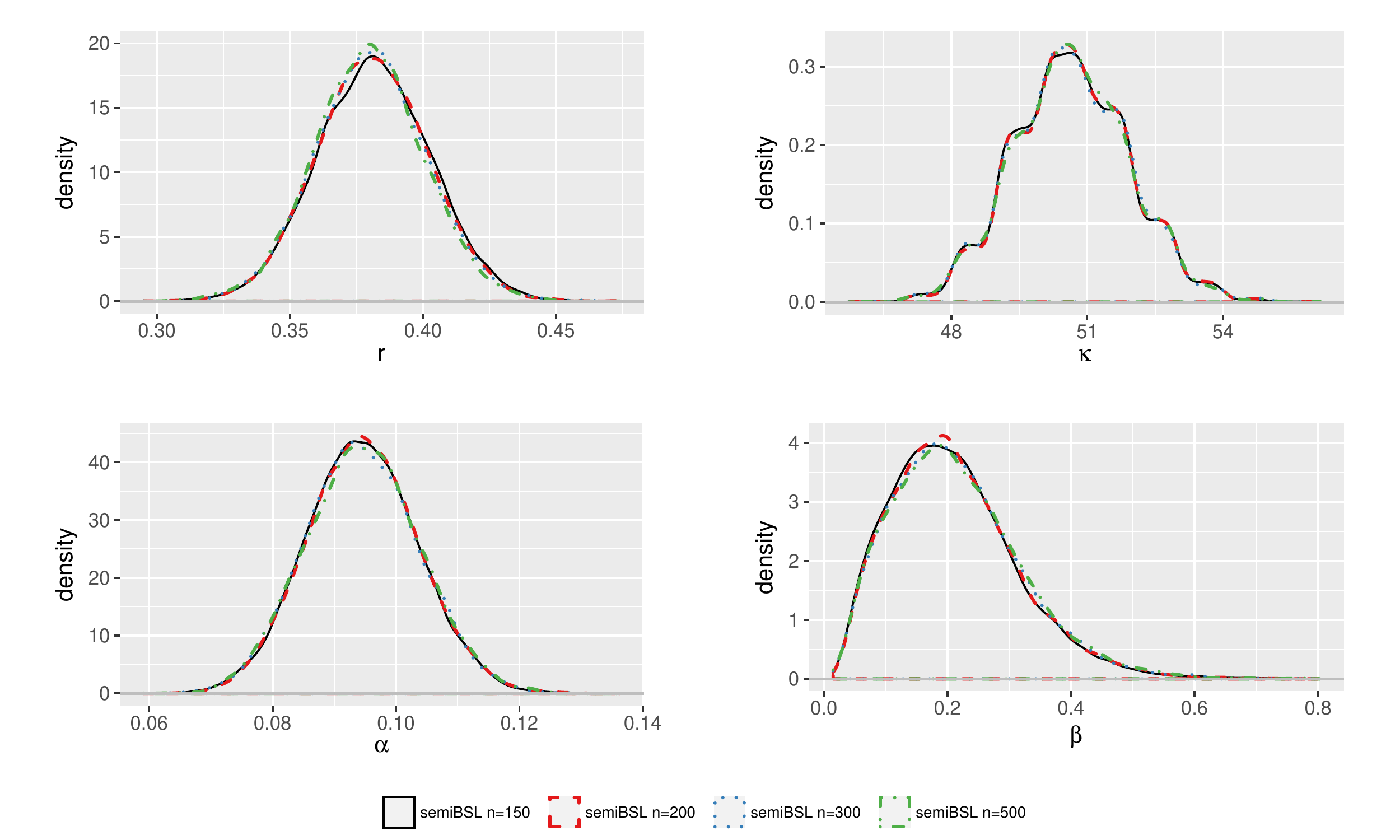}
			\caption{Approximate marginal posterior distributions for the boom and bust example with various number of simulations per iteration.}
			\label{fig:sens_bnb}
		\end{figure}

		Intuitively in the semiBSL likelihood function (\eqref{alg:pdf_semiBSL}), the value of $n$ is important to the approximation bias and variability of the marginal PDF and the correlation matrix. In this section, we show the semiBSL approximate marginal posteriors with different values of $n$ for each example (Figure \ref{fig:sens_ma2} - \ref{fig:sens_bnb}). With the minimum value of $n$ shown in the figure, we can still maintain a reasonable MCMC acceptance rate (eg $5\%$). It is apparent that the semiBSL approach is insensitive to $n$
		for all the examples.
		
		\newpage
		\section{SemiBSL with Warton's shrinkage estimation of correlation matrix} \label{app:semiBSL_warton}
		
		Here we demonstrate that is possible to use a shrinkage estimator for the correlation matrix of the Gaussian copula.  We illustrate this with the M/G/1 example. Table \ref{tab:prop_corr_mg1} shows the proportion of correlations and partial correlations below some certain thresholds (diagonal elements excluded). The ``true'' correlation matrix is calculated with $10,000$ independent simulations at the posterior means. It turns out that there is only small lag $1$ correlation and partial correlations for the chosen summary statistics.
		
		\begin{table}[!htp]
			\centering
			\begin{tabular}{ccccccc}
				\hline
				threshold & 0.001 & 0.002 & 0.005 & 0.01 & 0.05 & 0.1 \\
				\hline
				correlation & 0.05 & 0.12 & 0.31 & 0.57 & 0.97 & 1.00   \\
				partial correlation & 0.07 & 0.12 & 0.31 & 0.59 & 0.98 & 1.00 \\
				\hline
			\end{tabular}
			\caption{The proportion of correlations and partial correlations below certain thresholds in the ``true" correlation matrix of the summary statistics.}
			\label{tab:prop_corr_mg1}
		\end{table}
		
		Recall $\hat{\vect{R}}$ as an estimate of the correlation matrix. Then the Warton's correlation estimator \citep{Warton2008} is 
		
		\begin{equation*}
		\hat{\vect{R}}_{\lambda} = \lambda \hat{\vect{R}} + (1-\lambda) \vect{I}.
		\end{equation*}
		
\begin{figure}[!htp]
			\centering
			\includegraphics[width=0.95\textwidth]{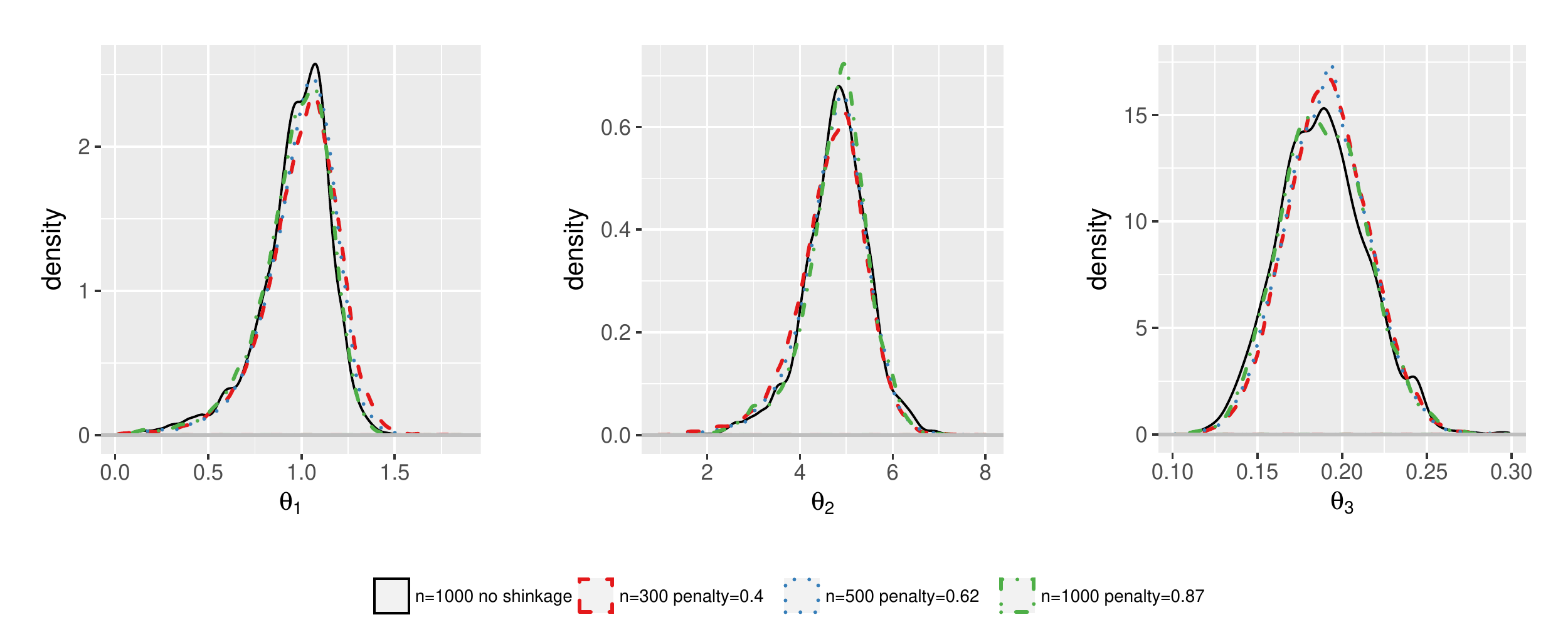}
			\caption{Approximate marginal posterior distributions for the M/G/1 example with Warton's shrinkage estimation.}
			\label{fig:post_warton_mg1}
		\end{figure}

		For a given $\lambda$, the shrinkage estimator is fast to compute. We adopt the same tuning approach as in \citet{An2018}. The value of $\lambda$ is selected so that the standard deviation of the semiBSL log-likelihood estimator is roughly $1.5$ at a parameter value with reasonable posterior support. The marginal posteriors with three different sets of $n$ and $\lambda$ are shown in Figure \ref{fig:post_warton_mg1}. With a penalty value of $0.4$, we managed to reduce $70\%$ simulations per iteration and still produce a close posterior to the result without shrinkage. The Warton's shrinkage estimator is very helpful in reducing the computational cost in this example, but the user should keep in mind that the performance is application dependent and should firstly check if a shrinkage estimator is likely to be appropriate.
		
	\end{appendices}

\end{document}